\documentclass[%
 reprint,
 amsmath,amssymb,
 aps,
]{revtex4-1}

\usepackage{xcolor}
\usepackage{graphicx}
\usepackage{dcolumn}
\usepackage{physics}
\usepackage{bm}

\usepackage{graphicx,color}
\usepackage{dcolumn}
\usepackage{bm}
\usepackage{amsmath,amssymb,mathrsfs}
\usepackage{url}
\usepackage{hyperref}

\usepackage[section]{placeins}
\usepackage{afterpage}

\newcommand{\C}{\mathbb{C}}

\newcommand{\set}[1]{\mathsf{#1}}
\newcommand{\grp}[1]{\mathsf{#1}}
\newcommand{\spc}[1]{\mathcal{#1}}

\def\d{{\rm d}}

\def\>{\rangle}
\def\<{\langle}
\def\\pi\pi{\>\!\>}

\newcommand{\map}[1]{\mathcal{#1}}

\newtheorem{defi}{Definition}

\begin{document}


\title{Quantum and Classical Data Transmission  Through Completely Depolarising Channels in a Superposition of Cyclic Orders}
\author{Giulio Chiribella}%
 \email{giulio@cs.hku.hk}
\affiliation{QICI Quantum Information and Computation Initiative, Department of Computer Science, The University of Hong Kong, Pokfulam Road, Hong Kong}
\affiliation{Department of Physics, The University of Hong Kong, Pokfulam Road, Hong Kong}
\affiliation{Department of Computer Science, University of Oxford, Wolfson Building, Parks Road, Oxford, United Kingdom}
\affiliation{HKU-Oxford Joint Laboratory for Quantum Information and Computation}  
\affiliation{Perimeter Institute for Theoretical Physics, 31 Caroline Street North, Waterloo, Ontario, Canada}
\author{Matt Wilson}
\affiliation{Department of Computer Science, University of Oxford, Wolfson Building, Parks Road, Oxford, United Kingdom}
\affiliation{HKU-Oxford Joint Laboratory for Quantum Information and Computation}  
\author{H. F. Chau}%
\affiliation{Department of Physics, The University of Hong Kong, Pokfulam Road, Hong Kong}




\date{\today}

\begin{abstract}
Completely depolarising channels are often regarded as the  prototype of physical processes  that are useless for communication:  any message that passes through them along a well-defined trajectory is completely erased. When two such channels are used in a quantum superposition of two alternative orders,  they become able to transmit some amount of classical information, but still no quantum information can pass through them. 
Here we show that the ability to place $N$ completely depolarising channels in a superposition of $N$ alternative causal orders 
enables a  high-fidelity, heralded transmission of quantum information with error vanishing as $1/N$. 
This phenomenon  highlights a fundamental difference with the $N=2$ case, where completely depolarising channels are unable to transmit quantum data, even when placed in a superposition of causal orders. The ability to place quantum channels in a superposition of orders also  leads to an increase of the classical communication capacity with $N$, which we   rigorously prove by deriving an exact  single-letter expression. Our results highlight the more complex patterns of  correlations arising from  multiple causal orders,  which is similar to the more complex patterns of entanglement arising in multipartite quantum systems. 
  \end{abstract}

\maketitle


{\em Introduction.}~Shannon's information theory was initially developed under the assumption that the information carriers were classical systems \cite{shannon1948mathematical}.  At the fundamental level, however,  physical systems obey the laws of quantum mechanics, which enable radically new communication protocols \cite{BEN84,Ekert1991QuantumTheorem} and give rise to a variety of new communication capacities \cite{wilde2013quantum}.  

Traditionally, the extension of Shannon's theory to the quantum domain assumed that the configuration of the communication  devices was fixed. In principle, however, quantum theory is compatible with scenarios where the communication devices are arranged in a coherent superposition of alternative  configurations.  For example, the available devices could act in different orders, and the choice of order could be controlled by the state of a quantum system, using a primitive known as the quantum {\tt SWITCH}  \cite{chiribella2009beyond,Chiribella2013}.  
  Similarly, the devices could be used as  alternatives to one another, and the choice of which device is  used for communication  could be controlled by the state of a quantum system, giving rise to a superposition of alternative quantum evolutions \cite{Aharonov1990,oi2003interference,gisin2005error,abbott2020communication,chiribella2019quantum,dong2019controlled}.  
   
The ability to superpose different configurations of communication devices can be exploited to achieve advantages over the standard model of quantum Shannon theory, where the configuration of the channels is fixed.   Advantages of the superposition of orders have been shown in   Refs. \cite{ebler2018enhanced, salek2018quantum, chiribella2021indefinite,procopio2019communication,procopio2020sending,loizeau2020channel,bhattacharya2021random}, while advantages of the superposition of channels have been shown in    Refs \cite{gisin2005error,abbott2020communication,chiribella2019quantum}.   At a conceptual level, these advantages can be rigorously formalised  in a resource-theoretic framework, where the resources are communication devices, and the allowed operations on them include  {placement operations}, which  determine the arrangement of the communication devices in space and time \cite{kristjansson2020resource}.  
Different  advantages  can  then be understood as the result of different ways to enlarge the set of  placement operations allowed by standard quantum Shannon theory.  At a more practical level, new communication  protocols with superpositions of configurations  have been experimentally realised \cite{lamoureux2005experimental,goswami2020increasing,guo2020experimental,rubino2021experimental,goswami2020experiments}. 
  The information-theoretic advantages of the superposition of causal orders have also inspired a new line of investigation in quantum thermodynamics  \cite{felce2020quantum}.
 
One of the most striking advantages of the superposition of configurations is the ability to communicate through channels that completely block information when used in a definite configuration.  The prototype of such channel is the completely depolarising channel, which outputs white noise independently of its input. Strikingly, it was shown that 
 two completely depolarising channels can be used  for  transmitting  classical information when arranged in a superposition of two alternative orders  \cite{ebler2018enhanced}. On the other hand, this phenomenon is limited to the transmission of classical bits:  in this Letter we will show that, when  two completely depolarising channels are combined in the quantum {\tt SWITCH}, the resulting channel cannot be used to transmit  quantum data. 

While the communication advantages of the quantum SWITCH of two channels are well known, much less is known about the advantages of the quantum SWITCH of multiple  channels.  Recent works  \cite{procopio2019communication,procopio2020sending}  considered the amount of classical bits transmitted through $N$ completely depolarising channels, showing an increase of the Holevo information  \cite{holevo1973bounds}. However, the Holevo information is only a lower bound to the actual capacity \cite{holevo1998capacity,schumacher1997sending}, and an increase in the Holevo information does not necessarily imply an increase in the capacity.  Moreover, the increase in the capacity, while technically interesting,  would only be a quantitative improvement in a task that can already be accomplished with $N=2$ channels. 
A natural question is whether there exists some  communication task  that cannot be achieved at all by superposing the order of two channels but instead becomes  possible when multiple channels are used.

Here, we answer the question in the affirmative, providing a concrete example of a communication task that can only achieved when $N>2$ causal orders are superposed. We consider  $N$ completely depolarising channels combined in a superposition of $N$ causal orders related to each other by cyclic permutations.
We show that a high-fidelity  heralded transmission of  quantum bits  can be achieved with error vanishing as $1/N$.  
 Our finding is  in stark contrast to the impossibility of quantum data transmission through  $N=2$ completely depolarising channels, and highlights a genuinely new feature arising from $N>2$  channels in  alternative causal orders. {The high-fidelity heralded transmission of quantum data is also potentially relevant for the task of entanglement distribution in quantum networks \cite{Bennett1993TeleportingChannels} 
 and for the task of private classical communication \cite{devetak2005private,horodecki2005secure}.  }

In addition to establishing the possibility of heralded quantum communication, we analytically determine the classical communication capacity of $N$ completely depolarising channels in a superposition of $N$ causal orders,  and we  demonstrate that the capacity increases monotonically with $N$.  
To this purpose,  we establish  a  connection between  the quantum {\tt SWITCH} of completely depolarising channels and the universal quantum {\tt NOT} gate \cite{bu2000universal,bu2000universal,ricci2004teleportation,de2004contextual,lim2011experimental}. We then use this connection to prove a single-letter formula for the classical capacity.  
 Our result demonstrates that increasing the number of ``useless'' channels leads to an increase in the number of   bits that can be reliably transmitted.    To the best of our knowledge, this is the first rigorous demonstration of a task where  the benefit of the superposition of causal orders grows monotonically  with the number of configurations that are superposed.

\begin{figure}[htbp!]
\centering
\includegraphics[scale = 0.215]{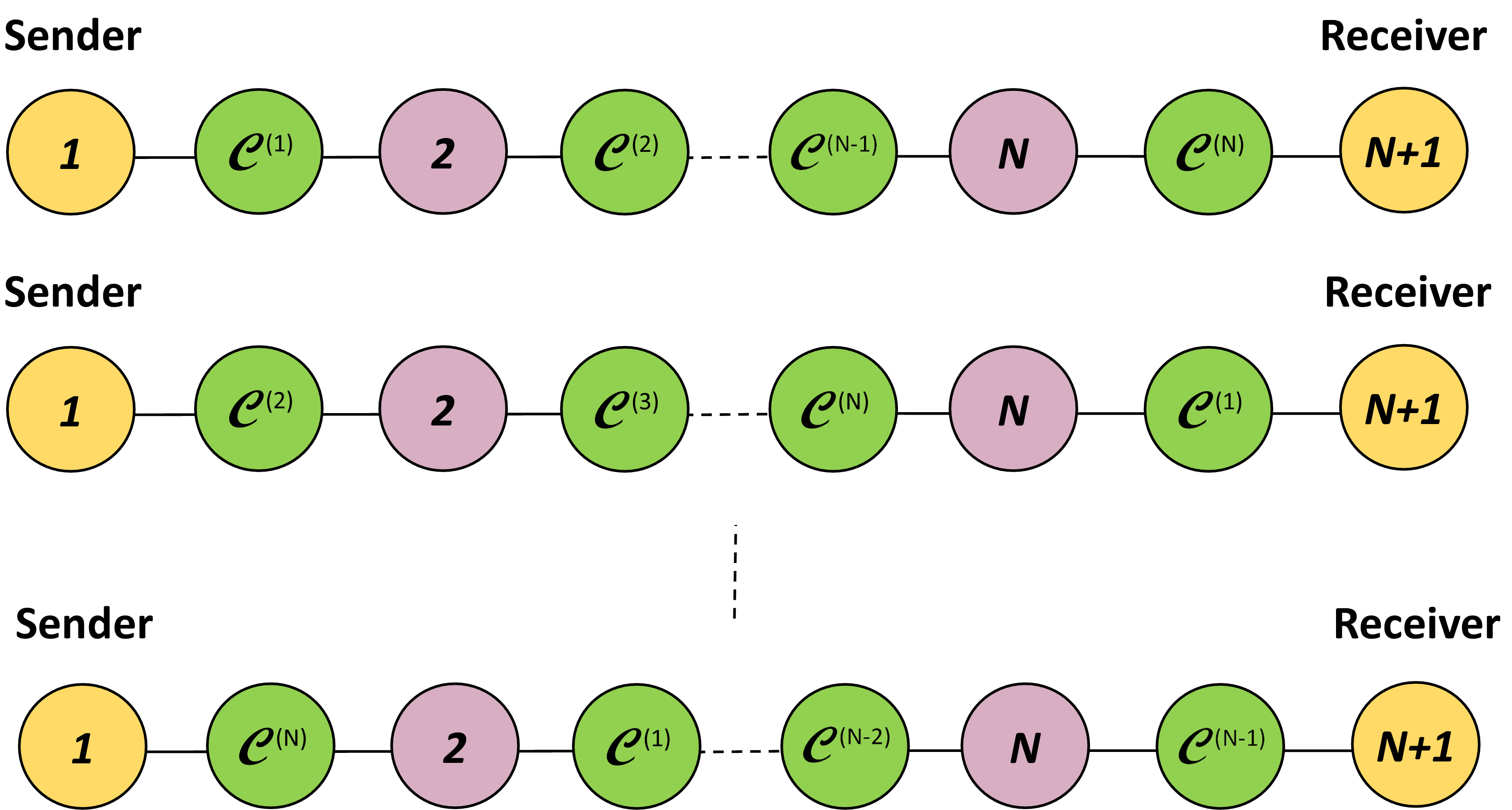}
\caption{{\bf Communication through  $N$ channels in a superposition of $N$ cyclic orders.} A sender, located at node 1 of a quantum communication network, sends messages to a receiver, located at node $N+1$, through a sequence of intermediate nodes, labelled as $2, \dots ,N$. The intermediate nodes are connected by $N$ quantum channels $\map C^{(1)}, \dots, \map C^{(N)}$, which have been placed in one of  $N$ configurations related by cyclic permutations, as shown in the graphic.  The choice of configuration is  controlled by a quantum system in a coherent superposition.}
\label{fig:network}
\end{figure}

\medskip

{\em Communication devices in a quantum superposition of alternative orders.}~A  communication device transmitting a quantum system   is described by a quantum channel, that is,  a completely positive trace preserving  linear map $\mathcal{C}$ transforming linear operators on the system's Hilbert space $\spc H$.  Any such   map can  be written in the Kraus representation $\mathcal{C}(\rho) = \sum_{i}C_{i}\rho C^{\dagger}_{i}$, where the Kraus operators $\{  C_{i} \}$ satisfy $ \sum_{i}C_{i}^{\dagger}C_{i} = I$.

Here, we consider the application of $N$  channels in a coherent superposition of different alternative orders. The superposition of orders is constructed using  the quantum {\tt SWITCH}  \cite{chiribella2009beyond,Chiribella2013}, a higher-order operation that transforms two quantum channels into a new quantum channel, in which the two input channels are executed in one of two alternative orders, depending on the state of  control qubit, called the order qubit.  Here we adopt  the original definition of the quantum {\tt SWITCH}  \cite{chiribella2009beyond}, where the two channels act in two subsequent time steps, possibly allowing for intermediate operations. 
Mathematically, the quantum {\tt SWITCH} transforms two input quantum channels $\mathcal{C}^{(1)}$ and $\mathcal{C}^{(2)}$   into the  output channel
\begin{align}\label{switch}
\map S  [\mathcal{C}^{(1)} ,  \mathcal{C}^{(2)}]  (\cdot)  =  \sum_{j_1,  j_2}    W_{j_1 j_2}  \cdot   W_{j_1j_2}^\dag  \, ,
\end{align}
whose  Kraus operators $W_{ij}$ are defined as 
\begin{align}
W_{ij}   :  =  |0\>\<0| \otimes C^{(1)}_{j_1}  \otimes  C^{(2)}_{j_2}      +|1\>\<1| \otimes     C^{(2)}_{j_2}  \otimes  C^{(1)}_{j_1}  \, ,     
\end{align}
where  the three systems in the tensor product on the right-hand side are the order qubit, the input system in the first time step, and the input system in the second time step. Here,  $\{ C^{(1)}_{j_1}\}$ and $\{   C^{(2)}_{j_2}\}$ are Kraus operators for channels $\map C^{(1)}$ and $\map C^{(2)}$, respectively. Note that, while the individual Kraus operators $W_{ij}$ depend on the choice of Kraus representation for $\map C^{(1)}$ and $\map C^{(2)}$,   the overall  quantum channel $\map S  [\mathcal{C}^{(1)} ,  \mathcal{C}^{(2)}]$ depends only on the channels $\map C^{(1)}$ and $\map C^{(2)}$ themselves, making the quantum {\tt SWITCH} a well-defined operation on quantum channels  \cite{chiribella2008transforming,Chiribella2013}.

 It is worth stressing that, while the order of the two processes $\mathcal{C}^{(1)}$ and $\mathcal{C}^{(2)}$ inside the quantum SWITCH is indefinite, the channel $ \map S  [\mathcal{C}^{(1)} ,  \mathcal{C}^{(2)}] $ produced by the quantum SWITCH has a well-defined causal structure:    the input of the first time slot is  provided first, followed by the output of the first time slot,   the input of the second time slot, and, finally,  the output of the second time slot. Accordingly, a communication protocol  using the channel $ \map S  [\mathcal{C}^{(1)} ,  \mathcal{C}^{(2)}] $   will have a well-defined causal structure: first, the sender inputs a state in the first time slot, then the first time slot is connected to the second with some intermediate operation, and finally the receiver collects the output of the second time slot. 

When $N> 2$ channels are available, the quantum {\tt SWITCH} operation (\ref{switch})  can be applied to each pair of channels, thus generating all possible permutations of their orders \cite{colnaghi2012quantum}.  In a resource theory of communication,  the quantum {\tt SWITCH} can be viewed as an operation performed by a communication provider, who places the available communication devices between the sender and receiver \cite{kristjansson2020resource}.   Here, we consider a placement of the $N$ devices in a network with $N-1$ intermediate nodes, as illustrated in Fig.~\ref{fig:network}.  
  Again, note that the causal structure of the process generated by the quantum SWITCH is well-defined, even though the $N$ channels inside the quantum SWITCH act in an indefinite order. As a consequence, the causal structure of the communication protocol in the network of Fig.~\ref{fig:network} is well-defined: first, the sender inputs the state in the first  node, then the first intermediate party receives the output at the second node and sends it to the third node, and so on  until, finally, the receiver receives the output at the last node. 

We  will assume that the order qubits are inaccessible to the sender and are initialised by the communication provider in a fixed state before the beginning of the communication protocol.   Also, we will take the intermediate nodes in Fig.~\ref{fig:network} to contain identity operations, so that the effective channel available to the sender and receiver becomes  
\begin{align}\label{effective}
     \map C_{{\rm eff}} (\rho)    &  = \sum_{\pi,  \pi' \in \set S}  \omega_{\pi ,\pi'} \ket{\pi}\bra{\pi'} \otimes \mathcal{C}_{\pi \pi'}  (\rho) \, ,
\end{align}
where   $\set S$ is a set of permutations,  $\omega$ is the state of the order qubits (with matrix elements $\omega_{\pi, \pi'}$ and support in a subspace spanned by  an orthonormal basis $\{|\pi\>\}_{\pi  \in \set S}$ labelled by permutations in $\set S$),  and 
\begin{align}
 \map C_{\pi \pi'}  (\rho) & :=\sum_{j_1,  \dots,  j_N} 
C^{\pi(1)  \cdots  \pi(N)}_{  j_{\pi(1)} ,\dots  j_{\pi  (N)}}
\, \rho \, C_{  j_{\pi'(1)} ,\dots  j_{\pi'  (N)}}^{\pi'(1) \cdots \pi'(N)   \,  \dag
}\end{align}  
with the notation $C^{i_1\cdots i_N}_{  j_{i_1}    \dots  j_{i_N}}   :  =  C^{(i_1)}_{j_{i_1}} \cdots  C^{(i_N)}_{j_{i_N}}$, where $\{C^{(i)}_{j_i}\}$ are Kraus operators for channel $\map C^{(i)}$.

\medskip

{\em Heralded quantum communication through completely depolarising channels.}~  
When the  configuration of the channels is fixed, the completely depolarising channel $\map D  (\cdot) :  =  I/d  \Tr[\cdot]$ is the prototype of a useless channel: since its  output  is independent of the input, this channel does not permit  the transmission of any data, be it classical or quantum.

Now, suppose that  $N$ completely depolarising channels are combined by the quantum {\tt SWITCH}, generating the effective channel   $\map C_{  {\rm eff}}$ in Equation (\ref{effective}).  
  In the following  we will take $\set S$ to be the set of cyclic permutations $\pi$, mapping the index $a$ into the index $\pi(a)   =(a   +  k)  \mod N  $ for some given $k\in \{0,\dots,  N-1\}$, and we will set  $\omega  =  |e_0\>\<e_0|$, with $|e_0\>    =  \sum_\pi   \, |\pi\>/\sqrt {N}$.   
  
      A convenient Kraus representation of the completely depolarising channel is  a uniform mixture of an orthogonal unitary basis $\{U_{i}\}_{i = 1}^{d^2}$, namely $\map D(\rho)=  \sum_{i=1}^{d^2} U_i \rho U_{i}^{\dagger}/d^2$, where $d$ is the dimension of the system and $\Tr [  U_i^\dag U_j  ]  =  d \, \delta_{i,j}$.   Using this representation, we  derive  the relations  
 \begin{align}
\label{aaa} 
\map C_{\pi \pi}    (\rho)  =  \frac I d    \qquad {\rm and} \qquad  \mathcal{C}_{\pi \pi'}  (\rho)= \frac{\rho}{d^2}  \quad \forall  \pi   \not =   \pi'\, , 
 \end{align}  
(see Appendix \ref{app:offdiag}). Inserting these relations  into Eq. (\ref{effective}) yields the expression
\begin{align}
  \nonumber  \map C_{{\rm eff}} (\rho)     & = \frac I N  \otimes \frac{I}{d} +
    \sum_{\pi \neq \pi'} \ket{\pi}\bra{\pi'} \otimes \frac{\rho}{Nd^2}   \\
   \label{aaaa}   &  =   \frac I N  \otimes \frac{I}{d} +
      \big( N  \,  |e_0\>\<e_0|  -  I  \big) \otimes \frac{\rho}{Nd^2}   \,    ,
    \end{align}
    the second equality following from the relations  $N |e_0\>\<e_0|=  \sum_{\pi,\pi'}   |\pi\>\<\pi'|$  and $I  = \sum_{\pi}  \,  |\pi\>\<\pi|$.
Rearranging the terms in Eq. (\ref{aaaa}), we rewrite the effective channel as 
\begin{align} 
   \map C_{{\rm eff}}   (\rho) & =  (1-p) \,   \rho_0   \otimes  \map E_0 (\rho)   +  p\,    \rho_1   \otimes \map E_1  (\rho)   \, , \label{heralded}
\end{align}
where   $\rho_0  :  = |e_0\>\<e_0|$ and $\rho_1:=    (I-|e_0\>\<e_0|)/(N-1)$ are  orthogonal states of the control system,  $\map E_0$ and $\map E_1$ are the quantum channels defined by
\begin{align}\label{E}
\map E_0 (\rho)   :=  \frac{  N-1 \,  }{  N-1 +  d^2}  \, \rho +   \frac{d^2}{  N-1 +  d^2} \,   \frac I d\, ,
\end{align}  
and 
\begin{align}\label{F}
\map E_1(\rho)   := \frac{  d^2   }{d^2-1}    \,   \frac I d    - \frac 1 {d^2-1} \, \rho
 \, , 
\end{align}  
 respectively, and  $p   :=  (N-1) (d^2-1)/(Nd^2)$.    Two  alternative ways to generate the channel $\map C_{\rm eff}$ from depolarising channels in a  superposition  cyclic orders  are discussed in Appendix  \ref{app:realisations}.

Equation (\ref{heralded}) shows that the effective channel $\map C_{\rm eff}$  is a mixture of two channels $\map E_0$ and $\map E_1$, flagged by two orthogonal states of the order qubits.  By measuring the order qubits, it is then possible to herald the occurrence of the channels $\map E_0$ and $\map E_1$. 

The channel $\map E_1$  is independent of $N$.  For $d=2$, it is  the universal {\tt NOT}  channel introduced by Bu\v zek, Hillery, and Werner \cite{bu2000universal} and experimentally realised in a series of works \cite{bu2000universal,ricci2004teleportation,de2004contextual,lim2011experimental}.  The universal {\tt NOT} gate is known to be an entanglement-breaking channel \cite{horodecki2003entanglement}, or equivalently, a ``measure-and-reprepare'' channel, which can be realised by measuring the input and preparing an output state depending on the measurement outcome \cite{werner1998optimal}.   Since entanglement-breaking channels have zero quantum capacity~\cite{holevo2001evaluating}, no quantum information can be transmitted through the channel $\map E_1$.     For $d>2$,  the channel $\map E_1$ is a generalisation of the universal {\tt NOT}, and can be characterised as the channel that minimises the fidelity between a generic input state $|\psi\>$ and the corresponding output state $\map E_1  (|\psi\>\<\psi|)$ (see Appendix \ref{app:unot}).  In Appendix \ref{app:unot}, we show  that $\map E_1$ is entanglement-breaking and therefore unable to transmit any quantum data.

  The channel $\map E_0$, instead,  is a depolarising channel, with probability of depolarisation equal to $d^2/(N  +  d^2-1)$.  Remarkably, this probability vanishes as $d^2/N$ in the large $N$ limit, enabling a perfect transmission of quantum data.     It is also remarkable that the probability of high-fidelity  transmission  does not vanish in the   large $N$ limit: such a probability remains  larger than $1/d^2$ for every  value of $N$. 
For qubits, this means that the state of the target system has a probability at least $25\%$ of reachhing the receiver with an error smaller than $4/N$.

{  The heralded,  high-fidelity  transmission of quantum information could be exploited  for the distribution of entanglement in quantum networks   \cite{Bennett1993TeleportingChannels}, which in turn serves as a primitive for distributed quantum computation \cite{buhrman2003distributed}.  Our results could also be useful  for cryptographic purposes, such as private classical communication \cite{devetak2005private,horodecki2005secure}, or the generation of  secret keys via the BB84 \cite{BEN84}  or E91 protocols \cite{Ekert1991QuantumTheorem}. A discussion of these  applications is provided  in Appendix \ref{app:applications}.     }

For finite $N$, it is possible to show that channel $\map E_0$ has a non-zero quantum capacity for all values of $N$ larger than a given finite value $N_0   > 2$.  For example, for  $d=2$ and $N > 13$ it is possible to show that  the probability of depolarisation is less than $1/4$,  which guarantees that the depolarising channel $\map E_0$ has a non-zero quantum capacity \cite{wilde2013quantum}.   In turn, the non-zero quantum capacity of channel $\map E_0$ ensures that the overall channel $\map C_{\rm eff}$ has a non-zero quantum capacity assisted by two-way classical communication \cite{bennett1996mixed}, as shown in  Appendix \ref{app:quantum}.   In Appendix  \ref{app:zerocap1}  we also show that quantum data transmission with the assistance of two-way classical communication is possible through the quantum SWITCH of $N$ cyclic permutations if and only if $N  \ge d+2$. 


The possibility of quantum information transmission is a fundamental difference between the bipartite and the multipartite quantum SWITCH in a way that  is somewhat reminiscent of the  difference between bipartite and multipartite entanglement.    For $N=2$, we prove that no superposition of causal orders
 permits the transmission of quantum bits through completely depolarising channels, under the natural assumption that the sender does not use the control system to establish entanglement with the receiver  (see Appendix \ref{app:zerocap} for the details). 
 
   {   All the results presented so far concerned the superposition of completely depolarising channels. A natural question is whether any of our conclusions would change if we were to consider partially depolarising channels.  In particular, one could ask  whether the  quantum SWITCH  could enable the transmission of quantum information using $N=2$ partially depolarising  channels that individually have  zero quantum capacity.   In Appendix \ref{app:partial},  we answer the question in the negative, showing that the quantum capacity of each depolarising channel is a bottleneck for the amount of quantum information one can send through the quantum SWITCH. An interesting open question is whether the use of partially depolarising channels could reduce  the number of channels  needed to achieve quantum data transmission starting from channels with zero capacity.  More broadly, the study of quantum communication with partially depolarising channels in an indefinite causal order remains as an  interesting problem for future research.}

{\em Enhanced transmission of classical information.}~We now quantify the amount of classical bits transmittable through $N$ depolarising channels in a superposition of $N$ alternative orders.  
By the Holevo-Schumacher-Westmoreland theorem  \cite{holevo1998capacity,schumacher1997sending}, the classical capacity of a generic noisy channel $\map N$ is given by  $C  (\map N)   =   \lim_{n\to \infty}    \chi (\map N^{\otimes n}) /n$, where $\chi $ is the Holevo information \cite{holevo1973bounds}, defined as  $\chi(\map N)  =  \max_{  (  \rho_x, p_x)_{x  \in \set X} }    S(\sum_x   p_x  \map N(\rho _x))    - \sum_x  p_x  \, S(\map C(\rho_x))$, $(\rho_x,  p_x)_{\in \set X}$ being an arbitrary ensemble of quantum states, and $S(\rho)   =  -\Tr[\rho \log \rho]$ being the von Neumann entropy.     In Appendix \ref{app:capswitch} we prove that the Holevo information of the effective channel $\map C_{\rm eff}$ is additive, and therefore the classical capacity has the single-letter formula $C(\map C_{\rm eff})    =  \chi (\map C_{\rm eff})$, for which we provide an exact expression.

\begin{figure}[h]
\centering
\includegraphics[width=0.5\textwidth]{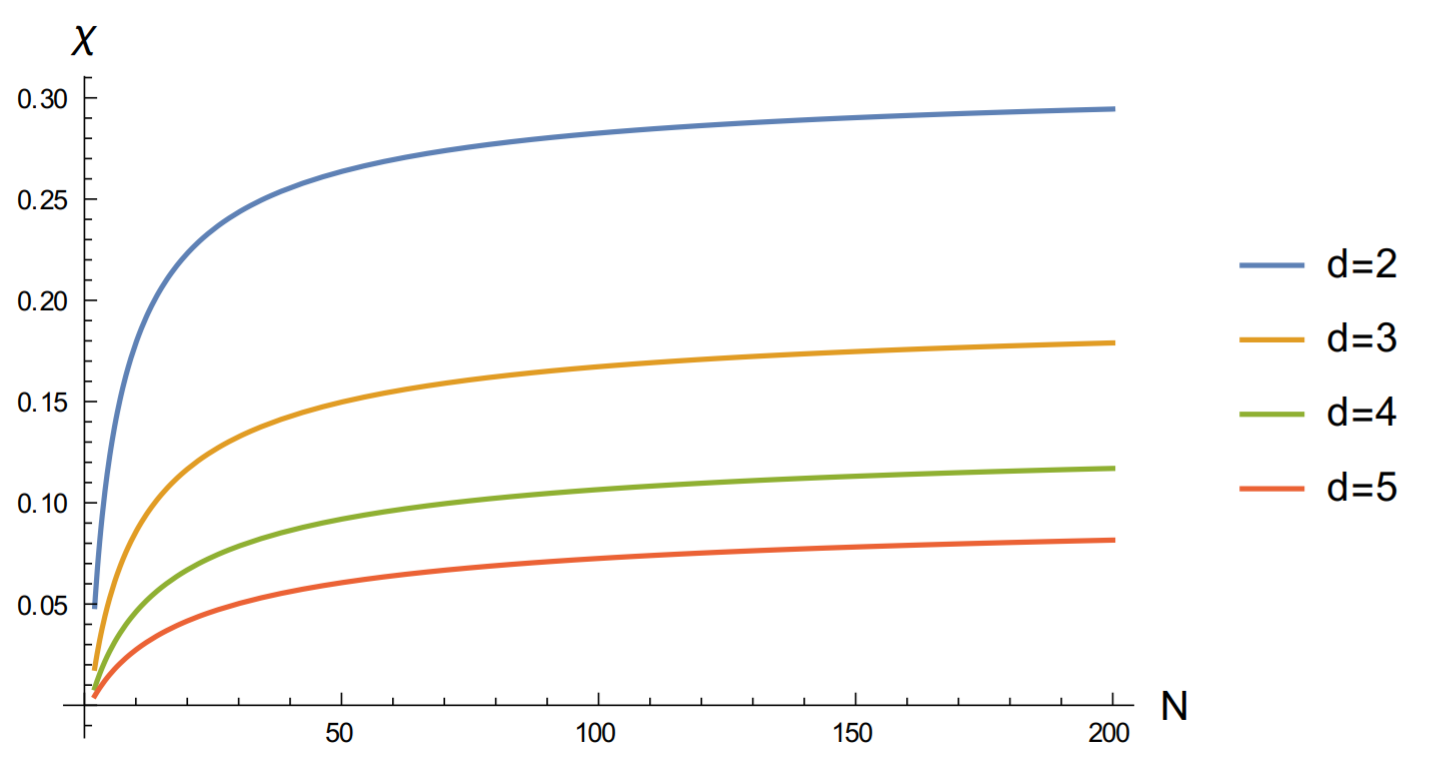}
\caption{ Classical capacity of the effective channel $\map C_{\rm eff}$,   plotted with respect to $N$ for message systems of dimension $d=2,3,4$ and $5$.}
\label{fig:plot}
\end{figure}

The classical capacity is plotted  in Fig. \ref{fig:plot} for different values of $N$ and $d$.
The capacity increases monotonically with $N$, rigorously demonstrating  the benefit of increasing the number of alternative orders.   In Appendix \ref{app:capswitch}   we provide an asymptotic expression for the capacity in the large $N$ limit, showing that it  decreases  with $d$, tending to zero for  $d\to \infty$.   For $N=2$, the decrease with $d$ was observed for the Holevo information \cite{ebler2018enhanced}, although it was not known whether the actual channel capacity was also decreasing.

{\em Conclusions.}~ We demonstrated a communication advantage of the superposition of multiple causal orders, by  showing a communication task that cannot be achieved by superposing two orders, but becomes possible when the number of orders is sufficiently large. Specifically, we demonstrated
 that the placement of $N$ completely depolarising channels in a superposition of  $N$ cyclic orders enables a high-fidelity heralded transmission of quantum information, with error  vanishing as $1/N$.   {For finite $N$,  we found that a non-zero quantum capacity assisted by two-way classical communication can be achieved with $N$ qubit depolarising channels whenever $N \ge 4$. } 
  
  
The possibility of quantum data transmission  through completely depolarising channels highlights a fundamental difference with the  $N=2$ scenario,   where no quantum information can pass through completely depolarising channels.  
A recent experiment \cite{taddei2021computational} on the superposition of $N=4$ channels  { suggests that an experimental demonstration of non-zero quantum capacity assisted by two-way classical communication  could be achieved  in the near future.}  Most importantly, our results  motivate an investigation of the operational features of the different types of quantum superpositions arising when multiple causal orders are superposed. 

It is intriguing to imagine that the  distinction between the superposition of two and multiple causal orders could mirror the distinction between bipartite and multipartite entanglement, whose study has led to the discovery of a wealth of new  quantum information protocols.  In this respect, our result indicates that the superposition of multiple causal orders is a genuinely new resource that is not reducible to the superposition of $N=2$ causal orders, just  as genuine multipartite entanglement is not reducible to bipartite entanglement.   We hope that our work will stimulate future explorations of the analogy between superpositions of causal orders and multipartite entanglement, thereby leading to a deeper understanding of the interplay between causality and quantum physics.   

\medskip 

{\em Acknowledgments.} We thank J Barrett, H Kristj\'ansson, and S Bhattacharya for helpful discussions.  This work was supported by the National Natural Science Foundation of China through grant 11675136, the Hong Kong Research Grant Council through grants 17300918 and 17307719  and through the Senior Research Fellowship Scheme SRFS2021-7S02, the Croucher Foundation,  the John Templeton Foundation through grant  61466, The Quantum Information Structure of Spacetime  (qiss.fr), the HKU Seed Funding for Basic Research, and the UK Engineering and Physical Sciences Research Council (EPSRC) through grant EP/L015242/1. 
Research at the Perimeter Institute is supported by the Government of Canada through the Department of Innovation, Science and Economic Development Canada and by the Province of Ontario through the Ministry of Research, Innovation and Science. The opinions expressed in this publication are those of the authors and do not necessarily reflect the views of the John Templeton Foundation.

\medskip 
{\em Note added.} After the completion of this work, we became aware of Ref. \cite{sazim2021classical}, which independently derived the Holevo information of $N$ completely depolarising channels in a superposition of  $N$ cyclic permutations.



\bibliography{references}

\begin{thebibliography}{65}%
\makeatletter
\providecommand \@ifxundefined [1]{%
 \@ifx{#1\undefined}
}%
\providecommand \@ifnum [1]{%
 \ifnum #1\expandafter \@firstoftwo
 \else \expandafter \@secondoftwo
 \fi
}%
\providecommand \@ifx [1]{%
 \ifx #1\expandafter \@firstoftwo
 \else \expandafter \@secondoftwo
 \fi
}%
\providecommand \natexlab [1]{#1}%
\providecommand \enquote  [1]{``#1''}%
\providecommand \bibnamefont  [1]{#1}%
\providecommand \bibfnamefont [1]{#1}%
\providecommand \citenamefont [1]{#1}%
\providecommand \href@noop [0]{\@secondoftwo}%
\providecommand \href [0]{\begingroup \@sanitize@url \@href}%
\providecommand \@href[1]{\@@startlink{#1}\@@href}%
\providecommand \@@href[1]{\endgroup#1\@@endlink}%
\providecommand \@sanitize@url [0]{\catcode `\\12\catcode `\$12\catcode
  `\&12\catcode `\#12\catcode `\^12\catcode `\_12\catcode `\%12\relax}%
\providecommand \@@startlink[1]{}%
\providecommand \@@endlink[0]{}%
\providecommand \url  [0]{\begingroup\@sanitize@url \@url }%
\providecommand \@url [1]{\endgroup\@href {#1}{\urlprefix }}%
\providecommand \urlprefix  [0]{URL }%
\providecommand \Eprint [0]{\href }%
\providecommand \doibase [0]{http://dx.doi.org/}%
\providecommand \selectlanguage [0]{\@gobble}%
\providecommand \bibinfo  [0]{\@secondoftwo}%
\providecommand \bibfield  [0]{\@secondoftwo}%
\providecommand \translation [1]{[#1]}%
\providecommand \BibitemOpen [0]{}%
\providecommand \bibitemStop [0]{}%
\providecommand \bibitemNoStop [0]{.\EOS\space}%
\providecommand \EOS [0]{\spacefactor3000\relax}%
\providecommand \BibitemShut  [1]{\csname bibitem#1\endcsname}%
\let\auto@bib@innerbib\@empty
\bibitem [{\citenamefont {Shannon}(1948)}]{shannon1948mathematical}%
  \BibitemOpen
  \bibfield  {author} {\bibinfo {author} {\bibfnamefont {C.~E.}\ \bibnamefont
  {Shannon}},\ }\href@noop {} {\bibfield  {journal} {\bibinfo  {journal} {The
  Bell System Technical Journal}\ }\textbf {\bibinfo {volume} {27}},\ \bibinfo
  {pages} {379} (\bibinfo {year} {1948})}\BibitemShut {NoStop}%
\bibitem [{\citenamefont {Bennett}\ and\ \citenamefont
  {Brassard}(1984)}]{BEN84}%
  \BibitemOpen
  \bibfield  {author} {\bibinfo {author} {\bibfnamefont {C.~H.}\ \bibnamefont
  {Bennett}}\ and\ \bibinfo {author} {\bibfnamefont {G.}~\bibnamefont
  {Brassard}},\ }in\ \href@noop {} {\emph {\bibinfo {booktitle} {Proceedings of
  IEEE International Conference on Computers, Systems, and Signal
  Processing}}}\ (\bibinfo {address} {India},\ \bibinfo {year} {1984})\ p.\
  \bibinfo {pages} {175}\BibitemShut {NoStop}%
\bibitem [{\citenamefont {Ekert}(1991)}]{Ekert1991QuantumTheorem}%
  \BibitemOpen
  \bibfield  {author} {\bibinfo {author} {\bibfnamefont {A.~K.}\ \bibnamefont
  {Ekert}},\ }\href {\doibase 10.1103/PhysRevLett.67.661} {\bibfield  {journal}
  {\bibinfo  {journal} {Physical Review Letters}\ }\textbf {\bibinfo {volume}
  {67}},\ \bibinfo {pages} {661} (\bibinfo {year} {1991})}\BibitemShut
  {NoStop}%
\bibitem [{\citenamefont {Wilde}(2013)}]{wilde2013quantum}%
  \BibitemOpen
  \bibfield  {author} {\bibinfo {author} {\bibfnamefont {M.~M.}\ \bibnamefont
  {Wilde}},\ }\href@noop {} {\emph {\bibinfo {title} {Quantum information
  theory}}}\ (\bibinfo  {publisher} {Cambridge University Press},\ \bibinfo
  {year} {2013})\BibitemShut {NoStop}%
\bibitem [{\citenamefont {Chiribella}\ \emph
  {et~al.}(2009{\natexlab{a}})\citenamefont {Chiribella}, \citenamefont
  {D’Ariano}, \citenamefont {Perinotti},\ and\ \citenamefont
  {Valiron}}]{chiribella2009beyond}%
  \BibitemOpen
  \bibfield  {author} {\bibinfo {author} {\bibfnamefont {G.}~\bibnamefont
  {Chiribella}}, \bibinfo {author} {\bibfnamefont {G.}~\bibnamefont
  {D’Ariano}}, \bibinfo {author} {\bibfnamefont {P.}~\bibnamefont
  {Perinotti}}, \ and\ \bibinfo {author} {\bibfnamefont {B.}~\bibnamefont
  {Valiron}},\ }\href@noop {} {\bibfield  {journal} {\bibinfo  {journal} {arXiv
  preprint arXiv:0912.0195}\ } (\bibinfo {year}
  {2009}{\natexlab{a}})}\BibitemShut {NoStop}%
\bibitem [{\citenamefont {Chiribella}\ \emph {et~al.}(2013)\citenamefont
  {Chiribella}, \citenamefont {D’Ariano}, \citenamefont {Perinotti},\ and\
  \citenamefont {Valiron}}]{Chiribella2013}%
  \BibitemOpen
  \bibfield  {author} {\bibinfo {author} {\bibfnamefont {G.}~\bibnamefont
  {Chiribella}}, \bibinfo {author} {\bibfnamefont {G.~M.}\ \bibnamefont
  {D’Ariano}}, \bibinfo {author} {\bibfnamefont {P.}~\bibnamefont
  {Perinotti}}, \ and\ \bibinfo {author} {\bibfnamefont {B.}~\bibnamefont
  {Valiron}},\ }\href@noop {} {\bibfield  {journal} {\bibinfo  {journal}
  {Physical Review A}\ }\textbf {\bibinfo {volume} {88}},\ \bibinfo {pages}
  {022318} (\bibinfo {year} {2013})}\BibitemShut {NoStop}%
\bibitem [{\citenamefont {Aharonov}\ \emph {et~al.}(1990)\citenamefont
  {Aharonov}, \citenamefont {Anandan}, \citenamefont {Popescu},\ and\
  \citenamefont {Vaidman}}]{Aharonov1990}%
  \BibitemOpen
  \bibfield  {author} {\bibinfo {author} {\bibfnamefont {Y.}~\bibnamefont
  {Aharonov}}, \bibinfo {author} {\bibfnamefont {J.}~\bibnamefont {Anandan}},
  \bibinfo {author} {\bibfnamefont {S.}~\bibnamefont {Popescu}}, \ and\
  \bibinfo {author} {\bibfnamefont {L.}~\bibnamefont {Vaidman}},\ }\href@noop
  {} {\bibfield  {journal} {\bibinfo  {journal} {Physical Review Letters}\
  }\textbf {\bibinfo {volume} {64}},\ \bibinfo {pages} {2965} (\bibinfo {year}
  {1990})}\BibitemShut {NoStop}%
\bibitem [{\citenamefont {Oi}(2003)}]{oi2003interference}%
  \BibitemOpen
  \bibfield  {author} {\bibinfo {author} {\bibfnamefont {D.~K.}\ \bibnamefont
  {Oi}},\ }\href@noop {} {\bibfield  {journal} {\bibinfo  {journal} {Physical
  Review Letters}\ }\textbf {\bibinfo {volume} {91}},\ \bibinfo {pages}
  {067902} (\bibinfo {year} {2003})}\BibitemShut {NoStop}%
\bibitem [{\citenamefont {Gisin}\ \emph {et~al.}(2005)\citenamefont {Gisin},
  \citenamefont {Linden}, \citenamefont {Massar},\ and\ \citenamefont
  {Popescu}}]{gisin2005error}%
  \BibitemOpen
  \bibfield  {author} {\bibinfo {author} {\bibfnamefont {N.}~\bibnamefont
  {Gisin}}, \bibinfo {author} {\bibfnamefont {N.}~\bibnamefont {Linden}},
  \bibinfo {author} {\bibfnamefont {S.}~\bibnamefont {Massar}}, \ and\ \bibinfo
  {author} {\bibfnamefont {S.}~\bibnamefont {Popescu}},\ }\href@noop {}
  {\bibfield  {journal} {\bibinfo  {journal} {Physical Review A}\ }\textbf
  {\bibinfo {volume} {72}},\ \bibinfo {pages} {012338} (\bibinfo {year}
  {2005})}\BibitemShut {NoStop}%
\bibitem [{\citenamefont {Abbott}\ \emph {et~al.}(2020)\citenamefont {Abbott},
  \citenamefont {Wechs}, \citenamefont {Horsman}, \citenamefont {Mhalla},\ and\
  \citenamefont {Branciard}}]{abbott2020communication}%
  \BibitemOpen
  \bibfield  {author} {\bibinfo {author} {\bibfnamefont {A.~A.}\ \bibnamefont
  {Abbott}}, \bibinfo {author} {\bibfnamefont {J.}~\bibnamefont {Wechs}},
  \bibinfo {author} {\bibfnamefont {D.}~\bibnamefont {Horsman}}, \bibinfo
  {author} {\bibfnamefont {M.}~\bibnamefont {Mhalla}}, \ and\ \bibinfo {author}
  {\bibfnamefont {C.}~\bibnamefont {Branciard}},\ }\href@noop {} {\bibfield
  {journal} {\bibinfo  {journal} {Quantum}\ }\textbf {\bibinfo {volume} {4}},\
  \bibinfo {pages} {333} (\bibinfo {year} {2020})}\BibitemShut {NoStop}%
\bibitem [{\citenamefont {Chiribella}\ and\ \citenamefont
  {Kristj{\'a}nsson}(2019)}]{chiribella2019quantum}%
  \BibitemOpen
  \bibfield  {author} {\bibinfo {author} {\bibfnamefont {G.}~\bibnamefont
  {Chiribella}}\ and\ \bibinfo {author} {\bibfnamefont {H.}~\bibnamefont
  {Kristj{\'a}nsson}},\ }\href@noop {} {\bibfield  {journal} {\bibinfo
  {journal} {Proceedings of the Royal Society A}\ }\textbf {\bibinfo {volume}
  {475}},\ \bibinfo {pages} {20180903} (\bibinfo {year} {2019})}\BibitemShut
  {NoStop}%
\bibitem [{\citenamefont {Dong}\ \emph {et~al.}(2019)\citenamefont {Dong},
  \citenamefont {Nakayama}, \citenamefont {Soeda},\ and\ \citenamefont
  {Murao}}]{dong2019controlled}%
  \BibitemOpen
  \bibfield  {author} {\bibinfo {author} {\bibfnamefont {Q.}~\bibnamefont
  {Dong}}, \bibinfo {author} {\bibfnamefont {S.}~\bibnamefont {Nakayama}},
  \bibinfo {author} {\bibfnamefont {A.}~\bibnamefont {Soeda}}, \ and\ \bibinfo
  {author} {\bibfnamefont {M.}~\bibnamefont {Murao}},\ }\href@noop {}
  {\bibfield  {journal} {\bibinfo  {journal} {arXiv preprint arXiv:1911.01645}\
  } (\bibinfo {year} {2019})}\BibitemShut {NoStop}%
\bibitem [{\citenamefont {Ebler}\ \emph {et~al.}(2018)\citenamefont {Ebler},
  \citenamefont {Salek},\ and\ \citenamefont {Chiribella}}]{ebler2018enhanced}%
  \BibitemOpen
  \bibfield  {author} {\bibinfo {author} {\bibfnamefont {D.}~\bibnamefont
  {Ebler}}, \bibinfo {author} {\bibfnamefont {S.}~\bibnamefont {Salek}}, \ and\
  \bibinfo {author} {\bibfnamefont {G.}~\bibnamefont {Chiribella}},\
  }\href@noop {} {\bibfield  {journal} {\bibinfo  {journal} {Physical Review
  Letters}\ }\textbf {\bibinfo {volume} {120}},\ \bibinfo {pages} {120502}
  (\bibinfo {year} {2018})}\BibitemShut {NoStop}%
\bibitem [{\citenamefont {Salek}\ \emph {et~al.}(2018)\citenamefont {Salek},
  \citenamefont {Ebler},\ and\ \citenamefont {Chiribella}}]{salek2018quantum}%
  \BibitemOpen
  \bibfield  {author} {\bibinfo {author} {\bibfnamefont {S.}~\bibnamefont
  {Salek}}, \bibinfo {author} {\bibfnamefont {D.}~\bibnamefont {Ebler}}, \ and\
  \bibinfo {author} {\bibfnamefont {G.}~\bibnamefont {Chiribella}},\
  }\href@noop {} {\bibfield  {journal} {\bibinfo  {journal} {arXiv preprint
  arXiv:1809.06655}\ } (\bibinfo {year} {2018})}\BibitemShut {NoStop}%
\bibitem [{\citenamefont {Chiribella}\ \emph {et~al.}(2021)\citenamefont
  {Chiribella}, \citenamefont {Banik}, \citenamefont {Bhattacharya},
  \citenamefont {Guha}, \citenamefont {Alimuddin}, \citenamefont {Roy},
  \citenamefont {Saha}, \citenamefont {Agrawal},\ and\ \citenamefont
  {Kar}}]{chiribella2021indefinite}%
  \BibitemOpen
  \bibfield  {author} {\bibinfo {author} {\bibfnamefont {G.}~\bibnamefont
  {Chiribella}}, \bibinfo {author} {\bibfnamefont {M.}~\bibnamefont {Banik}},
  \bibinfo {author} {\bibfnamefont {S.~S.}\ \bibnamefont {Bhattacharya}},
  \bibinfo {author} {\bibfnamefont {T.}~\bibnamefont {Guha}}, \bibinfo {author}
  {\bibfnamefont {M.}~\bibnamefont {Alimuddin}}, \bibinfo {author}
  {\bibfnamefont {A.}~\bibnamefont {Roy}}, \bibinfo {author} {\bibfnamefont
  {S.}~\bibnamefont {Saha}}, \bibinfo {author} {\bibfnamefont {S.}~\bibnamefont
  {Agrawal}}, \ and\ \bibinfo {author} {\bibfnamefont {G.}~\bibnamefont
  {Kar}},\ }\href@noop {} {\bibfield  {journal} {\bibinfo  {journal} {New
  Journal of Physics}\ }\textbf {\bibinfo {volume} {23}},\ \bibinfo {pages}
  {033039} (\bibinfo {year} {2021})}\BibitemShut {NoStop}%
\bibitem [{\citenamefont {Procopio}\ \emph {et~al.}(2019)\citenamefont
  {Procopio}, \citenamefont {Delgado}, \citenamefont {Enr{\'\i}quez},
  \citenamefont {Belabas},\ and\ \citenamefont
  {Levenson}}]{procopio2019communication}%
  \BibitemOpen
  \bibfield  {author} {\bibinfo {author} {\bibfnamefont {L.~M.}\ \bibnamefont
  {Procopio}}, \bibinfo {author} {\bibfnamefont {F.}~\bibnamefont {Delgado}},
  \bibinfo {author} {\bibfnamefont {M.}~\bibnamefont {Enr{\'\i}quez}}, \bibinfo
  {author} {\bibfnamefont {N.}~\bibnamefont {Belabas}}, \ and\ \bibinfo
  {author} {\bibfnamefont {J.~A.}\ \bibnamefont {Levenson}},\ }\href@noop {}
  {\bibfield  {journal} {\bibinfo  {journal} {Entropy}\ }\textbf {\bibinfo
  {volume} {21}},\ \bibinfo {pages} {1012} (\bibinfo {year}
  {2019})}\BibitemShut {NoStop}%
\bibitem [{\citenamefont {Procopio}\ \emph {et~al.}(2020)\citenamefont
  {Procopio}, \citenamefont {Delgado}, \citenamefont {Enr\'{\i}quez},
  \citenamefont {Belabas},\ and\ \citenamefont
  {Levenson}}]{procopio2020sending}%
  \BibitemOpen
  \bibfield  {author} {\bibinfo {author} {\bibfnamefont {L.~M.}\ \bibnamefont
  {Procopio}}, \bibinfo {author} {\bibfnamefont {F.}~\bibnamefont {Delgado}},
  \bibinfo {author} {\bibfnamefont {M.}~\bibnamefont {Enr\'{\i}quez}}, \bibinfo
  {author} {\bibfnamefont {N.}~\bibnamefont {Belabas}}, \ and\ \bibinfo
  {author} {\bibfnamefont {J.~A.}\ \bibnamefont {Levenson}},\ }\href {\doibase
  10.1103/PhysRevA.101.012346} {\bibfield  {journal} {\bibinfo  {journal}
  {Physical Review A}\ }\textbf {\bibinfo {volume} {101}},\ \bibinfo {pages}
  {012346} (\bibinfo {year} {2020})}\BibitemShut {NoStop}%
\bibitem [{\citenamefont {Loizeau}\ and\ \citenamefont
  {Grinbaum}(2020)}]{loizeau2020channel}%
  \BibitemOpen
  \bibfield  {author} {\bibinfo {author} {\bibfnamefont {N.}~\bibnamefont
  {Loizeau}}\ and\ \bibinfo {author} {\bibfnamefont {A.}~\bibnamefont
  {Grinbaum}},\ }\href@noop {} {\bibfield  {journal} {\bibinfo  {journal}
  {Physical Review A}\ }\textbf {\bibinfo {volume} {101}},\ \bibinfo {pages}
  {012340} (\bibinfo {year} {2020})}\BibitemShut {NoStop}%
\bibitem [{\citenamefont {Bhattacharya}\ \emph {et~al.}(2021)\citenamefont
  {Bhattacharya}, \citenamefont {Maity}, \citenamefont {Guha}, \citenamefont
  {Chiribella},\ and\ \citenamefont {Banik}}]{bhattacharya2021random}%
  \BibitemOpen
  \bibfield  {author} {\bibinfo {author} {\bibfnamefont {S.~S.}\ \bibnamefont
  {Bhattacharya}}, \bibinfo {author} {\bibfnamefont {A.~G.}\ \bibnamefont
  {Maity}}, \bibinfo {author} {\bibfnamefont {T.}~\bibnamefont {Guha}},
  \bibinfo {author} {\bibfnamefont {G.}~\bibnamefont {Chiribella}}, \ and\
  \bibinfo {author} {\bibfnamefont {M.}~\bibnamefont {Banik}},\ }\href
  {\doibase 10.1103/PRXQuantum.2.020350} {\bibfield  {journal} {\bibinfo
  {journal} {PRX Quantum}\ }\textbf {\bibinfo {volume} {2}},\ \bibinfo {pages}
  {020350} (\bibinfo {year} {2021})}\BibitemShut {NoStop}%
\bibitem [{\citenamefont {Kristjánsson}\ \emph {et~al.}(2020)\citenamefont
  {Kristjánsson}, \citenamefont {Chiribella}, \citenamefont {Salek},
  \citenamefont {Ebler},\ and\ \citenamefont
  {Wilson}}]{kristjansson2020resource}%
  \BibitemOpen
  \bibfield  {author} {\bibinfo {author} {\bibfnamefont {H.}~\bibnamefont
  {Kristjánsson}}, \bibinfo {author} {\bibfnamefont {G.}~\bibnamefont
  {Chiribella}}, \bibinfo {author} {\bibfnamefont {S.}~\bibnamefont {Salek}},
  \bibinfo {author} {\bibfnamefont {D.}~\bibnamefont {Ebler}}, \ and\ \bibinfo
  {author} {\bibfnamefont {M.}~\bibnamefont {Wilson}},\ }\href {\doibase
  10.1088/1367-2630/ab8ef7} {\bibfield  {journal} {\bibinfo  {journal} {New
  Journal of Physics}\ }\textbf {\bibinfo {volume} {22}},\ \bibinfo {pages}
  {073014} (\bibinfo {year} {2020})}\BibitemShut {NoStop}%
\bibitem [{\citenamefont {Lamoureux}\ \emph {et~al.}(2005)\citenamefont
  {Lamoureux}, \citenamefont {Brainis}, \citenamefont {Cerf}, \citenamefont
  {Emplit}, \citenamefont {Haelterman},\ and\ \citenamefont
  {Massar}}]{lamoureux2005experimental}%
  \BibitemOpen
  \bibfield  {author} {\bibinfo {author} {\bibfnamefont {L.-P.}\ \bibnamefont
  {Lamoureux}}, \bibinfo {author} {\bibfnamefont {E.}~\bibnamefont {Brainis}},
  \bibinfo {author} {\bibfnamefont {N.}~\bibnamefont {Cerf}}, \bibinfo {author}
  {\bibfnamefont {P.}~\bibnamefont {Emplit}}, \bibinfo {author} {\bibfnamefont
  {M.}~\bibnamefont {Haelterman}}, \ and\ \bibinfo {author} {\bibfnamefont
  {S.}~\bibnamefont {Massar}},\ }\href@noop {} {\bibfield  {journal} {\bibinfo
  {journal} {Physical Review Letters}\ }\textbf {\bibinfo {volume} {94}},\
  \bibinfo {pages} {230501} (\bibinfo {year} {2005})}\BibitemShut {NoStop}%
\bibitem [{\citenamefont {Goswami}\ \emph {et~al.}(2020)\citenamefont
  {Goswami}, \citenamefont {Cao}, \citenamefont {Paz-Silva}, \citenamefont
  {Romero},\ and\ \citenamefont {White}}]{goswami2020increasing}%
  \BibitemOpen
  \bibfield  {author} {\bibinfo {author} {\bibfnamefont {K.}~\bibnamefont
  {Goswami}}, \bibinfo {author} {\bibfnamefont {Y.}~\bibnamefont {Cao}},
  \bibinfo {author} {\bibfnamefont {G.}~\bibnamefont {Paz-Silva}}, \bibinfo
  {author} {\bibfnamefont {J.}~\bibnamefont {Romero}}, \ and\ \bibinfo {author}
  {\bibfnamefont {A.}~\bibnamefont {White}},\ }\href@noop {} {\bibfield
  {journal} {\bibinfo  {journal} {Physical Review Research}\ }\textbf {\bibinfo
  {volume} {2}},\ \bibinfo {pages} {033292} (\bibinfo {year}
  {2020})}\BibitemShut {NoStop}%
\bibitem [{\citenamefont {Guo}\ \emph {et~al.}(2020)\citenamefont {Guo},
  \citenamefont {Hu}, \citenamefont {Hou}, \citenamefont {Cao}, \citenamefont
  {Cui}, \citenamefont {Liu}, \citenamefont {Huang}, \citenamefont {Li},
  \citenamefont {Guo},\ and\ \citenamefont {Chiribella}}]{guo2020experimental}%
  \BibitemOpen
  \bibfield  {author} {\bibinfo {author} {\bibfnamefont {Y.}~\bibnamefont
  {Guo}}, \bibinfo {author} {\bibfnamefont {X.-M.}\ \bibnamefont {Hu}},
  \bibinfo {author} {\bibfnamefont {Z.-B.}\ \bibnamefont {Hou}}, \bibinfo
  {author} {\bibfnamefont {H.}~\bibnamefont {Cao}}, \bibinfo {author}
  {\bibfnamefont {J.-M.}\ \bibnamefont {Cui}}, \bibinfo {author} {\bibfnamefont
  {B.-H.}\ \bibnamefont {Liu}}, \bibinfo {author} {\bibfnamefont {Y.-F.}\
  \bibnamefont {Huang}}, \bibinfo {author} {\bibfnamefont {C.-F.}\ \bibnamefont
  {Li}}, \bibinfo {author} {\bibfnamefont {G.-C.}\ \bibnamefont {Guo}}, \ and\
  \bibinfo {author} {\bibfnamefont {G.}~\bibnamefont {Chiribella}},\
  }\href@noop {} {\bibfield  {journal} {\bibinfo  {journal} {Physical Review
  Letters}\ }\textbf {\bibinfo {volume} {124}},\ \bibinfo {pages} {030502}
  (\bibinfo {year} {2020})}\BibitemShut {NoStop}%
\bibitem [{\citenamefont {Rubino}\ \emph {et~al.}(2021)\citenamefont {Rubino},
  \citenamefont {Rozema}, \citenamefont {Ebler}, \citenamefont
  {Kristj{\'a}nsson}, \citenamefont {Salek}, \citenamefont {Gu{\'e}rin},
  \citenamefont {Abbott}, \citenamefont {Branciard}, \citenamefont {Brukner},
  \citenamefont {Chiribella},\ and\ \citenamefont
  {Walther}}]{rubino2021experimental}%
  \BibitemOpen
  \bibfield  {author} {\bibinfo {author} {\bibfnamefont {G.}~\bibnamefont
  {Rubino}}, \bibinfo {author} {\bibfnamefont {L.~A.}\ \bibnamefont {Rozema}},
  \bibinfo {author} {\bibfnamefont {D.}~\bibnamefont {Ebler}}, \bibinfo
  {author} {\bibfnamefont {H.}~\bibnamefont {Kristj{\'a}nsson}}, \bibinfo
  {author} {\bibfnamefont {S.}~\bibnamefont {Salek}}, \bibinfo {author}
  {\bibfnamefont {P.~A.}\ \bibnamefont {Gu{\'e}rin}}, \bibinfo {author}
  {\bibfnamefont {A.~A.}\ \bibnamefont {Abbott}}, \bibinfo {author}
  {\bibfnamefont {C.}~\bibnamefont {Branciard}}, \bibinfo {author}
  {\bibfnamefont {{\v{C}}.}~\bibnamefont {Brukner}}, \bibinfo {author}
  {\bibfnamefont {G.}~\bibnamefont {Chiribella}}, \ and\ \bibinfo {author}
  {\bibfnamefont {P.}~\bibnamefont {Walther}},\ }\href@noop {} {\bibfield
  {journal} {\bibinfo  {journal} {Physical Review Research}\ }\textbf {\bibinfo
  {volume} {3}},\ \bibinfo {pages} {013093} (\bibinfo {year}
  {2021})}\BibitemShut {NoStop}%
\bibitem [{\citenamefont {Goswami}\ and\ \citenamefont
  {Romero}(2020)}]{goswami2020experiments}%
  \BibitemOpen
  \bibfield  {author} {\bibinfo {author} {\bibfnamefont {K.}~\bibnamefont
  {Goswami}}\ and\ \bibinfo {author} {\bibfnamefont {J.}~\bibnamefont
  {Romero}},\ }\href@noop {} {\bibfield  {journal} {\bibinfo  {journal} {AVS
  Quantum Science}\ }\textbf {\bibinfo {volume} {2}},\ \bibinfo {pages}
  {037101} (\bibinfo {year} {2020})}\BibitemShut {NoStop}%
\bibitem [{\citenamefont {Felce}\ and\ \citenamefont
  {Vedral}(2020)}]{felce2020quantum}%
  \BibitemOpen
  \bibfield  {author} {\bibinfo {author} {\bibfnamefont {D.}~\bibnamefont
  {Felce}}\ and\ \bibinfo {author} {\bibfnamefont {V.}~\bibnamefont {Vedral}},\
  }\href@noop {} {\bibfield  {journal} {\bibinfo  {journal} {Physical review
  letters}\ }\textbf {\bibinfo {volume} {125}},\ \bibinfo {pages} {070603}
  (\bibinfo {year} {2020})}\BibitemShut {NoStop}%
\bibitem [{\citenamefont {Holevo}(1973)}]{holevo1973bounds}%
  \BibitemOpen
  \bibfield  {author} {\bibinfo {author} {\bibfnamefont {A.~S.}\ \bibnamefont
  {Holevo}},\ }\href@noop {} {\bibfield  {journal} {\bibinfo  {journal}
  {Problemy Peredachi Informatsii}\ }\textbf {\bibinfo {volume} {9}},\ \bibinfo
  {pages} {3} (\bibinfo {year} {1973})}\BibitemShut {NoStop}%
\bibitem [{\citenamefont {Holevo}(1998)}]{holevo1998capacity}%
  \BibitemOpen
  \bibfield  {author} {\bibinfo {author} {\bibfnamefont {A.~S.}\ \bibnamefont
  {Holevo}},\ }\href@noop {} {\bibfield  {journal} {\bibinfo  {journal} {IEEE
  Transactions on Information Theory}\ }\textbf {\bibinfo {volume} {44}},\
  \bibinfo {pages} {269} (\bibinfo {year} {1998})}\BibitemShut {NoStop}%
\bibitem [{\citenamefont {Schumacher}\ and\ \citenamefont
  {Westmoreland}(1997)}]{schumacher1997sending}%
  \BibitemOpen
  \bibfield  {author} {\bibinfo {author} {\bibfnamefont {B.}~\bibnamefont
  {Schumacher}}\ and\ \bibinfo {author} {\bibfnamefont {M.~D.}\ \bibnamefont
  {Westmoreland}},\ }\href@noop {} {\bibfield  {journal} {\bibinfo  {journal}
  {Physical Review A}\ }\textbf {\bibinfo {volume} {56}},\ \bibinfo {pages}
  {131} (\bibinfo {year} {1997})}\BibitemShut {NoStop}%
\bibitem [{\citenamefont {Bennett}\ \emph {et~al.}(1993)\citenamefont
  {Bennett}, \citenamefont {Brassard}, \citenamefont {Cr{\'{e}}peau},
  \citenamefont {Jozsa}, \citenamefont {Peres},\ and\ \citenamefont
  {Wootters}}]{Bennett1993TeleportingChannels}%
  \BibitemOpen
  \bibfield  {author} {\bibinfo {author} {\bibfnamefont {C.~H.}\ \bibnamefont
  {Bennett}}, \bibinfo {author} {\bibfnamefont {G.}~\bibnamefont {Brassard}},
  \bibinfo {author} {\bibfnamefont {C.}~\bibnamefont {Cr{\'{e}}peau}}, \bibinfo
  {author} {\bibfnamefont {R.}~\bibnamefont {Jozsa}}, \bibinfo {author}
  {\bibfnamefont {A.}~\bibnamefont {Peres}}, \ and\ \bibinfo {author}
  {\bibfnamefont {W.~K.}\ \bibnamefont {Wootters}},\ }\href {\doibase
  10.1103/PhysRevLett.70.1895} {\bibfield  {journal} {\bibinfo  {journal}
  {Physical Review Letters}\ }\textbf {\bibinfo {volume} {70}},\ \bibinfo
  {pages} {1895} (\bibinfo {year} {1993})}\BibitemShut {NoStop}%
\bibitem [{\citenamefont {Devetak}(2005)}]{devetak2005private}%
  \BibitemOpen
  \bibfield  {author} {\bibinfo {author} {\bibfnamefont {I.}~\bibnamefont
  {Devetak}},\ }\href@noop {} {\bibfield  {journal} {\bibinfo  {journal} {IEEE
  Transactions on Information Theory}\ }\textbf {\bibinfo {volume} {51}},\
  \bibinfo {pages} {44} (\bibinfo {year} {2005})}\BibitemShut {NoStop}%
\bibitem [{\citenamefont {Horodecki}\ \emph {et~al.}(2005)\citenamefont
  {Horodecki}, \citenamefont {Horodecki}, \citenamefont {Horodecki},\ and\
  \citenamefont {Oppenheim}}]{horodecki2005secure}%
  \BibitemOpen
  \bibfield  {author} {\bibinfo {author} {\bibfnamefont {K.}~\bibnamefont
  {Horodecki}}, \bibinfo {author} {\bibfnamefont {M.}~\bibnamefont
  {Horodecki}}, \bibinfo {author} {\bibfnamefont {P.}~\bibnamefont
  {Horodecki}}, \ and\ \bibinfo {author} {\bibfnamefont {J.}~\bibnamefont
  {Oppenheim}},\ }\href@noop {} {\bibfield  {journal} {\bibinfo  {journal}
  {Physical Review Letters}\ }\textbf {\bibinfo {volume} {94}},\ \bibinfo
  {pages} {160502} (\bibinfo {year} {2005})}\BibitemShut {NoStop}%
\bibitem [{\citenamefont {Bu{\v z}ek}\ \emph {et~al.}(2000)\citenamefont {Bu{\v
  z}ek}, \citenamefont {Hillery},\ and\ \citenamefont
  {Werner}}]{bu2000universal}%
  \BibitemOpen
  \bibfield  {author} {\bibinfo {author} {\bibfnamefont {V.}~\bibnamefont
  {Bu{\v z}ek}}, \bibinfo {author} {\bibfnamefont {M.}~\bibnamefont {Hillery}},
  \ and\ \bibinfo {author} {\bibfnamefont {F.}~\bibnamefont {Werner}},\
  }\href@noop {} {\bibfield  {journal} {\bibinfo  {journal} {Journal of Modern
  Optics}\ }\textbf {\bibinfo {volume} {47}},\ \bibinfo {pages} {211} (\bibinfo
  {year} {2000})}\BibitemShut {NoStop}%
\bibitem [{\citenamefont {Ricci}\ \emph {et~al.}(2004)\citenamefont {Ricci},
  \citenamefont {Sciarrino}, \citenamefont {Sias},\ and\ \citenamefont
  {De~Martini}}]{ricci2004teleportation}%
  \BibitemOpen
  \bibfield  {author} {\bibinfo {author} {\bibfnamefont {M.}~\bibnamefont
  {Ricci}}, \bibinfo {author} {\bibfnamefont {F.}~\bibnamefont {Sciarrino}},
  \bibinfo {author} {\bibfnamefont {C.}~\bibnamefont {Sias}}, \ and\ \bibinfo
  {author} {\bibfnamefont {F.}~\bibnamefont {De~Martini}},\ }\href@noop {}
  {\bibfield  {journal} {\bibinfo  {journal} {Physical Review Letters}\
  }\textbf {\bibinfo {volume} {92}},\ \bibinfo {pages} {047901} (\bibinfo
  {year} {2004})}\BibitemShut {NoStop}%
\bibitem [{\citenamefont {De~Martini}\ \emph {et~al.}(2004)\citenamefont
  {De~Martini}, \citenamefont {Pelliccia},\ and\ \citenamefont
  {Sciarrino}}]{de2004contextual}%
  \BibitemOpen
  \bibfield  {author} {\bibinfo {author} {\bibfnamefont {F.}~\bibnamefont
  {De~Martini}}, \bibinfo {author} {\bibfnamefont {D.}~\bibnamefont
  {Pelliccia}}, \ and\ \bibinfo {author} {\bibfnamefont {F.}~\bibnamefont
  {Sciarrino}},\ }\href@noop {} {\bibfield  {journal} {\bibinfo  {journal}
  {Physical Review Letters}\ }\textbf {\bibinfo {volume} {92}},\ \bibinfo
  {pages} {067901} (\bibinfo {year} {2004})}\BibitemShut {NoStop}%
\bibitem [{\citenamefont {Lim}\ \emph {et~al.}(2011)\citenamefont {Lim},
  \citenamefont {Kim}, \citenamefont {Ra}, \citenamefont {Bae},\ and\
  \citenamefont {Kim}}]{lim2011experimental}%
  \BibitemOpen
  \bibfield  {author} {\bibinfo {author} {\bibfnamefont {H.-T.}\ \bibnamefont
  {Lim}}, \bibinfo {author} {\bibfnamefont {Y.-S.}\ \bibnamefont {Kim}},
  \bibinfo {author} {\bibfnamefont {Y.-S.}\ \bibnamefont {Ra}}, \bibinfo
  {author} {\bibfnamefont {J.}~\bibnamefont {Bae}}, \ and\ \bibinfo {author}
  {\bibfnamefont {Y.-H.}\ \bibnamefont {Kim}},\ }\href@noop {} {\bibfield
  {journal} {\bibinfo  {journal} {Physical Review Letters}\ }\textbf {\bibinfo
  {volume} {107}},\ \bibinfo {pages} {160401} (\bibinfo {year}
  {2011})}\BibitemShut {NoStop}%
\bibitem [{\citenamefont {Chiribella}\ \emph {et~al.}(2008)\citenamefont
  {Chiribella}, \citenamefont {D'Ariano},\ and\ \citenamefont
  {Perinotti}}]{chiribella2008transforming}%
  \BibitemOpen
  \bibfield  {author} {\bibinfo {author} {\bibfnamefont {G.}~\bibnamefont
  {Chiribella}}, \bibinfo {author} {\bibfnamefont {G.~M.}\ \bibnamefont
  {D'Ariano}}, \ and\ \bibinfo {author} {\bibfnamefont {P.}~\bibnamefont
  {Perinotti}},\ }\href@noop {} {\bibfield  {journal} {\bibinfo  {journal} {EPL
  (Europhysics Letters)}\ }\textbf {\bibinfo {volume} {83}},\ \bibinfo {pages}
  {30004} (\bibinfo {year} {2008})}\BibitemShut {NoStop}%
\bibitem [{\citenamefont {Colnaghi}\ \emph {et~al.}(2012)\citenamefont
  {Colnaghi}, \citenamefont {Ariano}, \citenamefont {Facchini},\ and\
  \citenamefont {Perinotti}}]{colnaghi2012quantum}%
  \BibitemOpen
  \bibfield  {author} {\bibinfo {author} {\bibfnamefont {T.}~\bibnamefont
  {Colnaghi}}, \bibinfo {author} {\bibfnamefont {G.~M.}\ \bibnamefont
  {Ariano}}, \bibinfo {author} {\bibfnamefont {S.}~\bibnamefont {Facchini}}, \
  and\ \bibinfo {author} {\bibfnamefont {P.}~\bibnamefont {Perinotti}},\
  }\href@noop {} {\bibfield  {journal} {\bibinfo  {journal} {Physics Letters
  A}\ }\textbf {\bibinfo {volume} {376}},\ \bibinfo {pages} {2940} (\bibinfo
  {year} {2012})}\BibitemShut {NoStop}%
\bibitem [{\citenamefont {Horodecki}\ \emph {et~al.}(2003)\citenamefont
  {Horodecki}, \citenamefont {Shor},\ and\ \citenamefont
  {Ruskai}}]{horodecki2003entanglement}%
  \BibitemOpen
  \bibfield  {author} {\bibinfo {author} {\bibfnamefont {M.}~\bibnamefont
  {Horodecki}}, \bibinfo {author} {\bibfnamefont {P.~W.}\ \bibnamefont {Shor}},
  \ and\ \bibinfo {author} {\bibfnamefont {M.~B.}\ \bibnamefont {Ruskai}},\
  }\href@noop {} {\bibfield  {journal} {\bibinfo  {journal} {Reviews in
  Mathematical Physics}\ }\textbf {\bibinfo {volume} {15}},\ \bibinfo {pages}
  {629} (\bibinfo {year} {2003})}\BibitemShut {NoStop}%
\bibitem [{\citenamefont {Werner}(1998)}]{werner1998optimal}%
  \BibitemOpen
  \bibfield  {author} {\bibinfo {author} {\bibfnamefont {R.~F.}\ \bibnamefont
  {Werner}},\ }\href@noop {} {\bibfield  {journal} {\bibinfo  {journal}
  {Physical Review A}\ }\textbf {\bibinfo {volume} {58}},\ \bibinfo {pages}
  {1827} (\bibinfo {year} {1998})}\BibitemShut {NoStop}%
\bibitem [{\citenamefont {Holevo}\ and\ \citenamefont
  {Werner}(2001)}]{holevo2001evaluating}%
  \BibitemOpen
  \bibfield  {author} {\bibinfo {author} {\bibfnamefont {A.~S.}\ \bibnamefont
  {Holevo}}\ and\ \bibinfo {author} {\bibfnamefont {R.~F.}\ \bibnamefont
  {Werner}},\ }\href@noop {} {\bibfield  {journal} {\bibinfo  {journal}
  {Physical Review A}\ }\textbf {\bibinfo {volume} {63}},\ \bibinfo {pages}
  {032312} (\bibinfo {year} {2001})}\BibitemShut {NoStop}%
\bibitem [{\citenamefont {Buhrman}\ and\ \citenamefont
  {R{\"o}hrig}(2003)}]{buhrman2003distributed}%
  \BibitemOpen
  \bibfield  {author} {\bibinfo {author} {\bibfnamefont {H.}~\bibnamefont
  {Buhrman}}\ and\ \bibinfo {author} {\bibfnamefont {H.}~\bibnamefont
  {R{\"o}hrig}},\ }in\ \href@noop {} {\emph {\bibinfo {booktitle}
  {International Symposium on Mathematical Foundations of Computer Science}}}\
  (\bibinfo {organization} {Springer},\ \bibinfo {year} {2003})\ pp.\ \bibinfo
  {pages} {1--20}\BibitemShut {NoStop}%
\bibitem [{\citenamefont {Bennett}\ \emph {et~al.}(1996)\citenamefont
  {Bennett}, \citenamefont {DiVincenzo}, \citenamefont {Smolin},\ and\
  \citenamefont {Wootters}}]{bennett1996mixed}%
  \BibitemOpen
  \bibfield  {author} {\bibinfo {author} {\bibfnamefont {C.~H.}\ \bibnamefont
  {Bennett}}, \bibinfo {author} {\bibfnamefont {D.~P.}\ \bibnamefont
  {DiVincenzo}}, \bibinfo {author} {\bibfnamefont {J.~A.}\ \bibnamefont
  {Smolin}}, \ and\ \bibinfo {author} {\bibfnamefont {W.~K.}\ \bibnamefont
  {Wootters}},\ }\href@noop {} {\bibfield  {journal} {\bibinfo  {journal}
  {Physical Review A}\ }\textbf {\bibinfo {volume} {54}},\ \bibinfo {pages}
  {3824} (\bibinfo {year} {1996})}\BibitemShut {NoStop}%
\bibitem [{\citenamefont {Taddei}\ \emph {et~al.}(2021)\citenamefont {Taddei},
  \citenamefont {Cari{\~n}e}, \citenamefont {Mart{\'\i}nez}, \citenamefont
  {Garc{\'\i}a}, \citenamefont {Guerrero}, \citenamefont {Abbott},
  \citenamefont {Ara{\'u}jo}, \citenamefont {Branciard}, \citenamefont
  {G{\'o}mez}, \citenamefont {Walborn} \emph
  {et~al.}}]{taddei2021computational}%
  \BibitemOpen
  \bibfield  {author} {\bibinfo {author} {\bibfnamefont {M.~M.}\ \bibnamefont
  {Taddei}}, \bibinfo {author} {\bibfnamefont {J.}~\bibnamefont {Cari{\~n}e}},
  \bibinfo {author} {\bibfnamefont {D.}~\bibnamefont {Mart{\'\i}nez}}, \bibinfo
  {author} {\bibfnamefont {T.}~\bibnamefont {Garc{\'\i}a}}, \bibinfo {author}
  {\bibfnamefont {N.}~\bibnamefont {Guerrero}}, \bibinfo {author}
  {\bibfnamefont {A.~A.}\ \bibnamefont {Abbott}}, \bibinfo {author}
  {\bibfnamefont {M.}~\bibnamefont {Ara{\'u}jo}}, \bibinfo {author}
  {\bibfnamefont {C.}~\bibnamefont {Branciard}}, \bibinfo {author}
  {\bibfnamefont {E.~S.}\ \bibnamefont {G{\'o}mez}}, \bibinfo {author}
  {\bibfnamefont {S.~P.}\ \bibnamefont {Walborn}},  \emph {et~al.},\
  }\href@noop {} {\bibfield  {journal} {\bibinfo  {journal} {PRX Quantum}\
  }\textbf {\bibinfo {volume} {2}},\ \bibinfo {pages} {010320} (\bibinfo {year}
  {2021})}\BibitemShut {NoStop}%
\bibitem [{\citenamefont {Sazim}\ \emph {et~al.}(2021)\citenamefont {Sazim},
  \citenamefont {Sedlak}, \citenamefont {Singh},\ and\ \citenamefont
  {Pati}}]{sazim2021classical}%
  \BibitemOpen
  \bibfield  {author} {\bibinfo {author} {\bibfnamefont {S.}~\bibnamefont
  {Sazim}}, \bibinfo {author} {\bibfnamefont {M.}~\bibnamefont {Sedlak}},
  \bibinfo {author} {\bibfnamefont {K.}~\bibnamefont {Singh}}, \ and\ \bibinfo
  {author} {\bibfnamefont {A.~K.}\ \bibnamefont {Pati}},\ }\href@noop {}
  {\bibfield  {journal} {\bibinfo  {journal} {Physical Review A}\ }\textbf
  {\bibinfo {volume} {103}},\ \bibinfo {pages} {062610} (\bibinfo {year}
  {2021})}\BibitemShut {NoStop}%
\bibitem [{\citenamefont {Wilson}\ and\ \citenamefont
  {Chiribella}(2021)}]{wilson2020diagrammatic}%
  \BibitemOpen
  \bibfield  {author} {\bibinfo {author} {\bibfnamefont {M.}~\bibnamefont
  {Wilson}}\ and\ \bibinfo {author} {\bibfnamefont {G.}~\bibnamefont
  {Chiribella}},\ }in\ \href {\doibase 10.4204/EPTCS.340.17} {\emph {\bibinfo
  {booktitle} {{\rm Proceedings 17th International Conference on} Quantum
  Physics and Logic}}},\ \bibinfo {series} {Electronic Proceedings in
  Theoretical Computer Science}, Vol.\ \bibinfo {volume} {340},\ \bibinfo
  {editor} {edited by\ \bibinfo {editor} {\bibfnamefont {P.}~\bibnamefont
  {Arrighi}}, \bibinfo {editor} {\bibfnamefont {S.}~\bibnamefont {Mansfield}},
  \bibinfo {editor} {\bibfnamefont {P.}~\bibnamefont {Panangaden}}, \ and\
  \bibinfo {editor} {\bibfnamefont {B.}~\bibnamefont {Valiron}}}\ (\bibinfo
  {publisher} {Open Publishing Association},\ \bibinfo {year} {2021})\ pp.\
  \bibinfo {pages} {333--348}\BibitemShut {NoStop}%
\bibitem [{\citenamefont {Lloyd}\ \emph {et~al.}(2011)\citenamefont {Lloyd},
  \citenamefont {Maccone}, \citenamefont {Garcia-Patron}, \citenamefont
  {Giovannetti}, \citenamefont {Shikano}, \citenamefont {Pirandola},
  \citenamefont {Rozema}, \citenamefont {Darabi}, \citenamefont {Soudagar},
  \citenamefont {Shalm},\ and\ \citenamefont {Steinberg}}]{lloyd2011closed}%
  \BibitemOpen
  \bibfield  {author} {\bibinfo {author} {\bibfnamefont {S.}~\bibnamefont
  {Lloyd}}, \bibinfo {author} {\bibfnamefont {L.}~\bibnamefont {Maccone}},
  \bibinfo {author} {\bibfnamefont {R.}~\bibnamefont {Garcia-Patron}}, \bibinfo
  {author} {\bibfnamefont {V.}~\bibnamefont {Giovannetti}}, \bibinfo {author}
  {\bibfnamefont {Y.}~\bibnamefont {Shikano}}, \bibinfo {author} {\bibfnamefont
  {S.}~\bibnamefont {Pirandola}}, \bibinfo {author} {\bibfnamefont {L.~A.}\
  \bibnamefont {Rozema}}, \bibinfo {author} {\bibfnamefont {A.}~\bibnamefont
  {Darabi}}, \bibinfo {author} {\bibfnamefont {Y.}~\bibnamefont {Soudagar}},
  \bibinfo {author} {\bibfnamefont {L.~K.}\ \bibnamefont {Shalm}}, \ and\
  \bibinfo {author} {\bibfnamefont {A.~M.}\ \bibnamefont {Steinberg}},\
  }\href@noop {} {\bibfield  {journal} {\bibinfo  {journal} {Physical Review
  Letters}\ }\textbf {\bibinfo {volume} {106}},\ \bibinfo {pages} {040403}
  (\bibinfo {year} {2011})}\BibitemShut {NoStop}%
\bibitem [{\citenamefont {Oeckl}(2008)}]{oeckl2008general}%
  \BibitemOpen
  \bibfield  {author} {\bibinfo {author} {\bibfnamefont {R.}~\bibnamefont
  {Oeckl}},\ }\href@noop {} {\bibfield  {journal} {\bibinfo  {journal}
  {Advances in Theoretical and Mathematical Physics}\ }\textbf {\bibinfo
  {volume} {12}},\ \bibinfo {pages} {319} (\bibinfo {year} {2008})}\BibitemShut
  {NoStop}%
\bibitem [{\citenamefont {Svetlichny}(2011)}]{svetlichny2011time}%
  \BibitemOpen
  \bibfield  {author} {\bibinfo {author} {\bibfnamefont {G.}~\bibnamefont
  {Svetlichny}},\ }\href@noop {} {\bibfield  {journal} {\bibinfo  {journal}
  {International Journal of Theoretical Physics}\ }\textbf {\bibinfo {volume}
  {50}},\ \bibinfo {pages} {3903} (\bibinfo {year} {2011})}\BibitemShut
  {NoStop}%
\bibitem [{\citenamefont {Oreshkov}\ and\ \citenamefont
  {Cerf}(2015)}]{oreshkov2015operational}%
  \BibitemOpen
  \bibfield  {author} {\bibinfo {author} {\bibfnamefont {O.}~\bibnamefont
  {Oreshkov}}\ and\ \bibinfo {author} {\bibfnamefont {N.~J.}\ \bibnamefont
  {Cerf}},\ }\href@noop {} {\bibfield  {journal} {\bibinfo  {journal} {Nature
  Physics}\ }\textbf {\bibinfo {volume} {11}},\ \bibinfo {pages} {853}
  (\bibinfo {year} {2015})}\BibitemShut {NoStop}%
\bibitem [{\citenamefont {Chiribella}\ \emph {et~al.}(2005)\citenamefont
  {Chiribella}, \citenamefont {D’Ariano}, \citenamefont {Perinotti},\ and\
  \citenamefont {Cerf}}]{chiribella2005extremal}%
  \BibitemOpen
  \bibfield  {author} {\bibinfo {author} {\bibfnamefont {G.}~\bibnamefont
  {Chiribella}}, \bibinfo {author} {\bibfnamefont {G.}~\bibnamefont
  {D’Ariano}}, \bibinfo {author} {\bibfnamefont {P.}~\bibnamefont
  {Perinotti}}, \ and\ \bibinfo {author} {\bibfnamefont {N.}~\bibnamefont
  {Cerf}},\ }\href@noop {} {\bibfield  {journal} {\bibinfo  {journal} {Physical
  Review A}\ }\textbf {\bibinfo {volume} {72}},\ \bibinfo {pages} {042336}
  (\bibinfo {year} {2005})}\BibitemShut {NoStop}%
\bibitem [{\citenamefont {Eisert}\ \emph {et~al.}(2020)\citenamefont {Eisert},
  \citenamefont {Hangleiter}, \citenamefont {Walk}, \citenamefont {Roth},
  \citenamefont {Markham}, \citenamefont {Parekh}, \citenamefont {Chabaud},\
  and\ \citenamefont {Kashefi}}]{eisert2020quantum}%
  \BibitemOpen
  \bibfield  {author} {\bibinfo {author} {\bibfnamefont {J.}~\bibnamefont
  {Eisert}}, \bibinfo {author} {\bibfnamefont {D.}~\bibnamefont {Hangleiter}},
  \bibinfo {author} {\bibfnamefont {N.}~\bibnamefont {Walk}}, \bibinfo {author}
  {\bibfnamefont {I.}~\bibnamefont {Roth}}, \bibinfo {author} {\bibfnamefont
  {D.}~\bibnamefont {Markham}}, \bibinfo {author} {\bibfnamefont
  {R.}~\bibnamefont {Parekh}}, \bibinfo {author} {\bibfnamefont
  {U.}~\bibnamefont {Chabaud}}, \ and\ \bibinfo {author} {\bibfnamefont
  {E.}~\bibnamefont {Kashefi}},\ }\href@noop {} {\bibfield  {journal} {\bibinfo
   {journal} {Nature Reviews Physics}\ }\textbf {\bibinfo {volume} {2}},\
  \bibinfo {pages} {382} (\bibinfo {year} {2020})}\BibitemShut {NoStop}%
\bibitem [{\citenamefont {Wu}\ and\ \citenamefont
  {Sanders}(2019)}]{wu2019efficient}%
  \BibitemOpen
  \bibfield  {author} {\bibinfo {author} {\bibfnamefont {Y.-D.}\ \bibnamefont
  {Wu}}\ and\ \bibinfo {author} {\bibfnamefont {B.~C.}\ \bibnamefont
  {Sanders}},\ }\href@noop {} {\bibfield  {journal} {\bibinfo  {journal} {New
  Journal of Physics}\ }\textbf {\bibinfo {volume} {21}},\ \bibinfo {pages}
  {073026} (\bibinfo {year} {2019})}\BibitemShut {NoStop}%
\bibitem [{\citenamefont {Wu}\ \emph {et~al.}(2021)\citenamefont {Wu},
  \citenamefont {Bai}, \citenamefont {Chiribella},\ and\ \citenamefont
  {Liu}}]{wu2021efficient}%
  \BibitemOpen
  \bibfield  {author} {\bibinfo {author} {\bibfnamefont {Y.-D.}\ \bibnamefont
  {Wu}}, \bibinfo {author} {\bibfnamefont {G.}~\bibnamefont {Bai}}, \bibinfo
  {author} {\bibfnamefont {G.}~\bibnamefont {Chiribella}}, \ and\ \bibinfo
  {author} {\bibfnamefont {N.}~\bibnamefont {Liu}},\ }\href@noop {} {\bibfield
  {journal} {\bibinfo  {journal} {Physical Review Letters}\ }\textbf {\bibinfo
  {volume} {126}},\ \bibinfo {pages} {240503} (\bibinfo {year}
  {2021})}\BibitemShut {NoStop}%
\bibitem [{\citenamefont {Christandl}\ and\ \citenamefont
  {Renner}(2012)}]{christandl2012reliable}%
  \BibitemOpen
  \bibfield  {author} {\bibinfo {author} {\bibfnamefont {M.}~\bibnamefont
  {Christandl}}\ and\ \bibinfo {author} {\bibfnamefont {R.}~\bibnamefont
  {Renner}},\ }\href@noop {} {\bibfield  {journal} {\bibinfo  {journal}
  {Physical Review Letters}\ }\textbf {\bibinfo {volume} {109}},\ \bibinfo
  {pages} {120403} (\bibinfo {year} {2012})}\BibitemShut {NoStop}%
\bibitem [{\citenamefont {Zhu}\ and\ \citenamefont
  {Hayashi}(2019)}]{zhu2019optimal}%
  \BibitemOpen
  \bibfield  {author} {\bibinfo {author} {\bibfnamefont {H.}~\bibnamefont
  {Zhu}}\ and\ \bibinfo {author} {\bibfnamefont {M.}~\bibnamefont {Hayashi}},\
  }\href@noop {} {\bibfield  {journal} {\bibinfo  {journal} {Physical Review
  A}\ }\textbf {\bibinfo {volume} {99}},\ \bibinfo {pages} {052346} (\bibinfo
  {year} {2019})}\BibitemShut {NoStop}%
\bibitem [{\citenamefont {Oreshkov}\ \emph {et~al.}(2012)\citenamefont
  {Oreshkov}, \citenamefont {Costa},\ and\ \citenamefont
  {Brukner}}]{oreshkov2012quantum}%
  \BibitemOpen
  \bibfield  {author} {\bibinfo {author} {\bibfnamefont {O.}~\bibnamefont
  {Oreshkov}}, \bibinfo {author} {\bibfnamefont {F.}~\bibnamefont {Costa}}, \
  and\ \bibinfo {author} {\bibfnamefont {{\v{C}}.}~\bibnamefont {Brukner}},\
  }\href@noop {} {\bibfield  {journal} {\bibinfo  {journal} {Nature
  Communications}\ }\textbf {\bibinfo {volume} {3}},\ \bibinfo {pages} {1092}
  (\bibinfo {year} {2012})}\BibitemShut {NoStop}%
\bibitem [{\citenamefont {Paw{\l}owski}\ and\ \citenamefont
  {Brunner}(2011)}]{pawlowski2011semi}%
  \BibitemOpen
  \bibfield  {author} {\bibinfo {author} {\bibfnamefont {M.}~\bibnamefont
  {Paw{\l}owski}}\ and\ \bibinfo {author} {\bibfnamefont {N.}~\bibnamefont
  {Brunner}},\ }\href@noop {} {\bibfield  {journal} {\bibinfo  {journal}
  {Physical Review A}\ }\textbf {\bibinfo {volume} {84}},\ \bibinfo {pages}
  {010302} (\bibinfo {year} {2011})}\BibitemShut {NoStop}%
\bibitem [{\citenamefont {Chaturvedi}\ \emph {et~al.}(2018)\citenamefont
  {Chaturvedi}, \citenamefont {Ray}, \citenamefont {Veynar},\ and\
  \citenamefont {Paw{\l}owski}}]{chaturvedi2018security}%
  \BibitemOpen
  \bibfield  {author} {\bibinfo {author} {\bibfnamefont {A.}~\bibnamefont
  {Chaturvedi}}, \bibinfo {author} {\bibfnamefont {M.}~\bibnamefont {Ray}},
  \bibinfo {author} {\bibfnamefont {R.}~\bibnamefont {Veynar}}, \ and\ \bibinfo
  {author} {\bibfnamefont {M.}~\bibnamefont {Paw{\l}owski}},\ }\href@noop {}
  {\bibfield  {journal} {\bibinfo  {journal} {Quantum Information Processing}\
  }\textbf {\bibinfo {volume} {17}},\ \bibinfo {pages} {131} (\bibinfo {year}
  {2018})}\BibitemShut {NoStop}%
\bibitem [{\citenamefont {Chiribella}\ \emph
  {et~al.}(2009{\natexlab{b}})\citenamefont {Chiribella}, \citenamefont
  {D’Ariano},\ and\ \citenamefont {Perinotti}}]{chiribella2009realization}%
  \BibitemOpen
  \bibfield  {author} {\bibinfo {author} {\bibfnamefont {G.}~\bibnamefont
  {Chiribella}}, \bibinfo {author} {\bibfnamefont {G.~M.}\ \bibnamefont
  {D’Ariano}}, \ and\ \bibinfo {author} {\bibfnamefont {P.}~\bibnamefont
  {Perinotti}},\ }\href@noop {} {\bibfield  {journal} {\bibinfo  {journal}
  {Journal of Mathematical physics}\ }\textbf {\bibinfo {volume} {50}},\
  \bibinfo {pages} {042101} (\bibinfo {year} {2009}{\natexlab{b}})}\BibitemShut
  {NoStop}%
\bibitem [{\citenamefont {Horodecki}\ and\ \citenamefont
  {Horodecki}(1999)}]{horodecki1999reduction}%
  \BibitemOpen
  \bibfield  {author} {\bibinfo {author} {\bibfnamefont {M.}~\bibnamefont
  {Horodecki}}\ and\ \bibinfo {author} {\bibfnamefont {P.}~\bibnamefont
  {Horodecki}},\ }\href@noop {} {\bibfield  {journal} {\bibinfo  {journal}
  {Physical Review A}\ }\textbf {\bibinfo {volume} {59}},\ \bibinfo {pages}
  {4206} (\bibinfo {year} {1999})}\BibitemShut {NoStop}%
\bibitem [{\citenamefont {Braunstein}\ \emph {et~al.}(1999)\citenamefont
  {Braunstein}, \citenamefont {Caves}, \citenamefont {Jozsa}, \citenamefont
  {Linden}, \citenamefont {Popescu},\ and\ \citenamefont
  {Schack}}]{braunstein1999separability}%
  \BibitemOpen
  \bibfield  {author} {\bibinfo {author} {\bibfnamefont {S.~L.}\ \bibnamefont
  {Braunstein}}, \bibinfo {author} {\bibfnamefont {C.~M.}\ \bibnamefont
  {Caves}}, \bibinfo {author} {\bibfnamefont {R.}~\bibnamefont {Jozsa}},
  \bibinfo {author} {\bibfnamefont {N.}~\bibnamefont {Linden}}, \bibinfo
  {author} {\bibfnamefont {S.}~\bibnamefont {Popescu}}, \ and\ \bibinfo
  {author} {\bibfnamefont {R.}~\bibnamefont {Schack}},\ }\href@noop {}
  {\bibfield  {journal} {\bibinfo  {journal} {Physical Review Letters}\
  }\textbf {\bibinfo {volume} {83}},\ \bibinfo {pages} {1054} (\bibinfo {year}
  {1999})}\BibitemShut {NoStop}%
\bibitem [{\citenamefont {Holevo}(2002)}]{holevo2002remarks}%
  \BibitemOpen
  \bibfield  {author} {\bibinfo {author} {\bibfnamefont {A.~S.}\ \bibnamefont
  {Holevo}},\ }\href@noop {} {\bibfield  {journal} {\bibinfo  {journal} {arXiv
  preprint quant-ph/0212025}\ } (\bibinfo {year} {2002})}\BibitemShut {NoStop}%
\bibitem [{\citenamefont {King}(2003)}]{king2003capacity}%
  \BibitemOpen
  \bibfield  {author} {\bibinfo {author} {\bibfnamefont {C.}~\bibnamefont
  {King}},\ }\href@noop {} {\bibfield  {journal} {\bibinfo  {journal} {IEEE
  Transactions on Information Theory}\ }\textbf {\bibinfo {volume} {49}},\
  \bibinfo {pages} {221} (\bibinfo {year} {2003})}\BibitemShut {NoStop}%
\bibitem [{\citenamefont {Shor}(2002)}]{shor2002additivity}%
  \BibitemOpen
  \bibfield  {author} {\bibinfo {author} {\bibfnamefont {P.~W.}\ \bibnamefont
  {Shor}},\ }\href@noop {} {\bibfield  {journal} {\bibinfo  {journal} {Journal
  of Mathematical Physics}\ }\textbf {\bibinfo {volume} {43}},\ \bibinfo
  {pages} {4334} (\bibinfo {year} {2002})}\BibitemShut {NoStop}%
\end{thebibliography}%

\appendix

\begin{widetext}  

\section{Off-diagonal terms of the cyclic switch}\label{app:offdiag}
We consider the terms for which $\pi \neq \pi'$ in
\begin{align}
     \map C_{\rm eff} (\rho)    &  = \sum_{\pi,  \pi' \in \set S}  \omega_{\pi ,\pi'} \ket{\pi}\bra{\pi'} \otimes \mathcal{C}_{\pi \pi'}  (\rho) \, ,
\end{align}
where given the Kraus decomposition $\{\frac{1}{d} U^{\pi (i)}_{j_{\pi(i)}}\}_{j_{\pi(i)}= 1}^{d^2}$ of each completely depolarising channel $\map D^{(i)}$
\begin{equation}
\map C_{\pi \pi'}  =\sum_{j_1,\dots,  j_N}  \frac{  U^{\pi (1)}_{j_{\pi(1)}} \dots U^{\pi(N)}_{j_{\pi(N)}} 
\,  \rho  \, U^{\pi'(N)\dagger}_{j_{\pi'(N)}} \dots  U^{\pi'(1)\dagger}_{j_{\pi'(1)} }
}{d^{2N}} .
\end{equation}
We show that any $\mathcal{C}_{\pi \pi'}$,  with $\pi \neq \pi'$ cyclic permutations,  evaluates to the same expression 
\begin{align}\mathcal{C}_{\pi \pi'} = \frac{\rho}{d^2} . 
\end{align}
The derivation relies on the property of the Kraus decomposition of a completely depolarising channel used in \cite{ebler2018enhanced} to derive the output of the 2 party switch of two completely depolarising channels.  Specifically, they form an orthonormal unitary basis on the $d^2$ dimensional space of linear operators where the orthonormality is with respect to the Hilbert Schmidt product. Since $\pi$ and $\pi'$ are cyclic permutations, $\pi'$ itself is a cyclic permutation relative to $\pi$. Each $\mathcal{C}_{\pi \pi'}$ can be evaluated using only the knowledge that the relative permutation between $\pi$ and $\pi'$ is a non-trivial cyclic permutation.
Adopting an ordered product notation
\begin{align}
\prod_{a = \pi(1)}^{\pi(N)}  U^{a}_{j_{a}} \equiv  U^{\pi (1)}_{j_{\pi(1)}} \dots U^{\pi(N)}_{j_{\pi(N)}} ,  \end{align}
the output can be written as
\begin{align}\mathcal{C}_{\pi \pi'} = \frac{1}{d^{2N}}  \sum^{d^2}_{\{j_{k}| k \in N \}} \Bigg( \prod_{a = \pi (1)}^{\pi (N)}  U^{a}_{j_{a}} \Bigg) \rho  \Bigg( \prod_{b = \pi'(N)}^{\pi'(1)} U^{b\dagger}_{j_{b}} \Bigg) . \end{align}
Using the cyclic property of $\pi'$ on the right hand side we can rewrite this as
\begin{align}\mathcal{C}_{\pi\pi'} = \frac{1}{d^{2N}}  \sum^{d^2}_{\{j_{k}| k \in N \}} \Bigg( \prod_{a = \pi (1)}^{\pi (N-1)}  U^{a}_{j_{a}} \Bigg) U^{\pi(N)}_{j_{\pi(N)}} \rho \Bigg(\prod_{b = \pi'(N)}^{\pi(1)}U^{b\dagger}_{j_{b}} \Bigg) U^{\pi(N)\dagger}_{j_{\pi(N)}} \Bigg( \prod_{c = \pi(N-1)}^{\pi'(1)} U^{c\dagger}_{j_{c}} \Bigg) \end{align}
Next, summing over $j_{\pi(N)}$ through the use of the identity
\begin{align}\label{iduno}\frac{1}{d^2}\sum^{d^2}_{j_{\pi(N)}=1} U_{j_{\pi(N)}}^{\pi({N})} \rho U_{j_{\pi(N)}}^{\pi({N}) \dagger} =  \Tr [\rho] \frac{I}{d} \end{align}
gives
\begin{align}
 \nonumber    \mathcal{C}_{\pi\pi'} & = \frac{d}{d^{2N}} \sum^{d^2}_{\{j_{k}| k \in N, k \neq \pi(N) \}} \Bigg( \prod_{a = \pi (1)}^{\pi (N-1)}  U^{a}_{j_{a}} \Bigg) \Tr [\rho \prod_{b = \pi'(N)}^{\pi(1)}U^{b\dagger}_{j_{b}}]\Bigg( \prod_{c = \pi(N-1)}^{\pi'(1)} U^{c\dagger}_{j_{c}} \Bigg) \\
     & = \frac{d}{d^{2N}} \sum^{d^2}_{\{j_{k}| k \in N, k \neq \pi(N) \}} \Bigg( \prod_{a = \pi (1)}^{\pi' (N)}  U^{a}_{j_{a}} \Bigg) \Tr [\rho \prod_{b = \pi'(N)}^{\pi(1)}U^{b\dagger}_{j_{b}}]  .
\end{align}
Then summing over $j_{\pi'(N)}$ through the use of the identity
\begin{align}\label{idue}\sum^{d^2}_{j_{\pi'(N)} = 1}U_{j_{\pi'(N)}} \Tr [\rho U^{\dagger}_{j_{\pi'(N)}}] = \rho d\end{align}
gives
\begin{align}
 \nonumber      \mathcal{C}_{\pi\pi'} = & \frac{1}{d^{2(N-1)}} \sum^{d^2}_{\{j_{k}| k \in N, k \neq \pi(N),\pi'(N) \}}  \prod_{a = \pi (1)}^{\pi' (N) - 1}  U^{a}_{j_{a}} \prod_{b = \pi'(N) - 1}^{\pi(1)}U^{b\dagger}_{j_{b}} \rho \\
     \nonumber  = & \frac{1}{d^{2(N-1)}} \sum^{d^2}_{\{j_{k}| k \in N, k \neq \pi(N),\pi'(N) \}}  \rho \\
      = & \frac{\rho}{d^2} .
\end{align}

The above relations can also be derived by diagrammatic means, as done in Ref. \cite{wilson2020diagrammatic}, where the case of arbitrary non-cyclic permutations is also analysed. Notably, cyclic permutations appear to be those for which the off-diagonal terms in $\map C_{\pi\pi'}$ have the  highest weight.

\section{Two realisations of the effective channel $\map C_{\rm eff}$}\label{app:realisations}

Here we discuss two alternative scenarios that give rise to  the superposition of $N$ depolarising channels corresponding to channel $\map C_{\rm eff}$ in Eq. (7) of the main text.     For simplicity of presentation, we will focus on the realisation of the depolarising channels as random mixtures of unitary processes, although all the arguments below can be extended to arbitrary realisations.

\begin{figure}[htbp!]
\centering
\includegraphics[scale = 0.4]{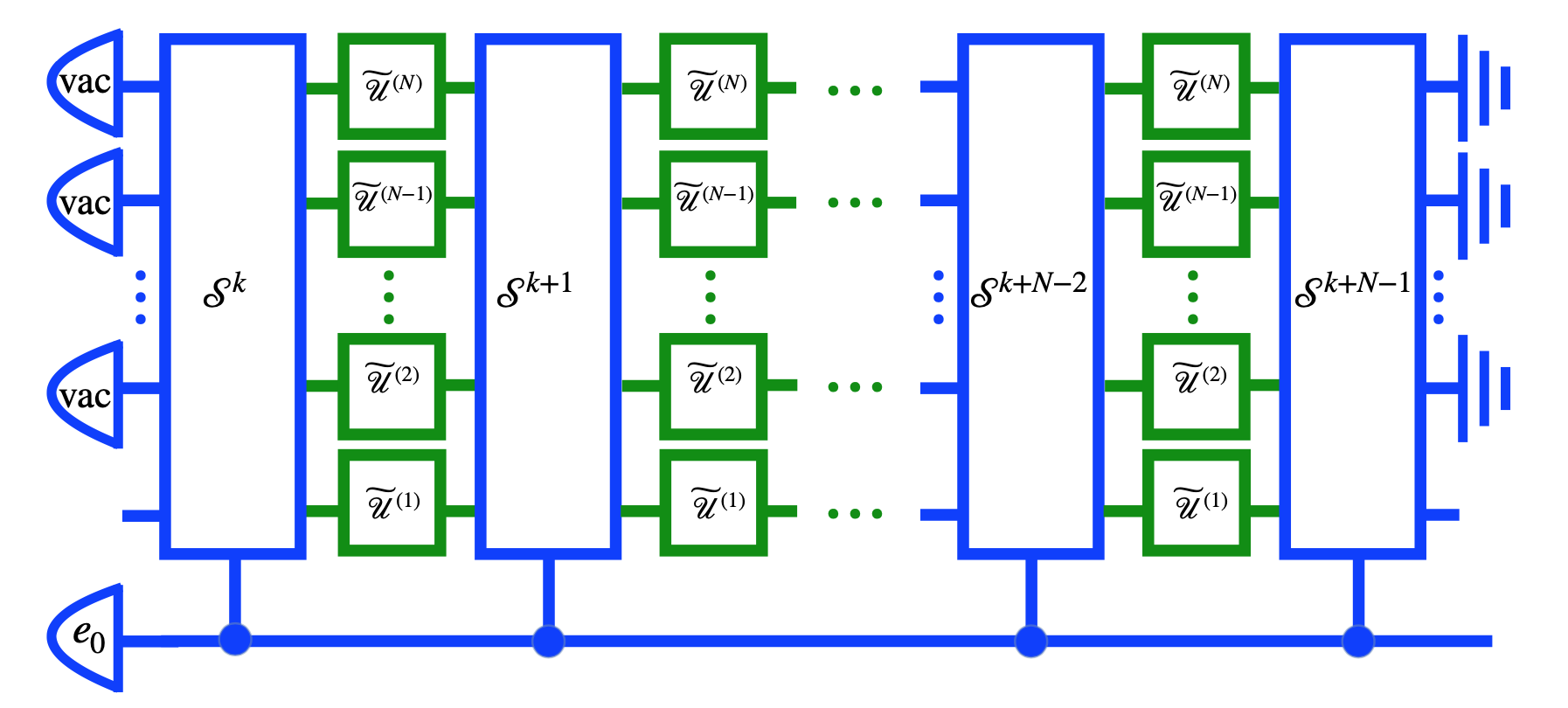}
\caption{{\bf Realisation of the effective channel $\map C_{\rm eff}$  through coherent control of paths.}   A quantum system traverses $N$ regions at $N$  distinct moments of time, following a path controlled by an $N$-dimensional control system (bottom wire).   The control system is initialised in the uniform superposition state $|e_0\>  =  \sum_{k=0}^{N-1}  |k\>/\sqrt{N}$, and  the control of the paths is implemented by unitary channels ${\rm ctrl}-\map S^{k+i}$, corresponding to unitary operators ${\rm ctrl}-S^{k+i}:  =  \sum_{k=0}^{N-1}    S^{k+i}  \otimes  |k\>\<k|$, where $S$ is the unitary operator that shifts cyclically the $N$ inputs by one position.   In the $k$-th region, the system undergoes a unitary process $\map U^{(k)}$, chosen at random according to a suitable probability  distribution. In the figure,  $\widetilde{\map U}^{(k)}$ is a unitary channel that extends $\map U^{(k)}$ to the $d+ 1$ space spanned by the states of the system and by an orthogonal vacuum state $|{\rm vac} \>$.   In the end, $N-1$ systems are discarded, and the overall evolution from the input to the output is given by the channel $\map C_{\rm eff}$ in Eq. (7) of the main text. }
\label{fig:paths}
\end{figure}

In the first scenario, illustrated in Figure~\ref{fig:paths},   a quantum system travels through $N$ regions (in green) of space at $N$ different moments of time.   When the system passes through the $k$-th region, it undergoes some random unitary process $\map U^{(k)}$, distributed according to a probability distribution such that the average process is completely depolarising. When the system does  not pass through the $k$-th region, the input to that region is the vacuum state $|{\rm vac}\>$.  
 Overall, the process in the region is described by a unitary channel $\widetilde{\map U}^{(k)}$, acting on the direct sum of a one-particle subspace and of the vacuum, and coinciding with $\map U^{(k)}$ on the states of the one-particle subspace \cite{chiribella2019quantum}.    
The path of the system through the $N$ regions is controlled by a quantum system (bottom wire in the figure), which permutes the system with the vacuum state of $N-1$ modes. 
  The overall evolution resulting from this scheme coincides with  the effective channel $\map C_{\rm eff}$ in Eq. (7) of the main text.  This way of reproducing the output of  the quantum SWITCH   was introduced in  \cite{chiribella2019quantum} for $N=2$ and is generalised here to $N>2$. 

It is important to note that this  realisation of the channel $\map C_{\rm eff}$ requires the unitary process $\widetilde{\map U}^{(k)}$  to be the same at all  time steps.    This situation can  be engineered  in a photonic table-top scenario, where the unitary processes are implemented by optical devices, such as waveplates, whose behaviour is stable over the timescale of the experiment. To reproduce the completely depolarising channel, the choice of waveplates is randomised, or simply taken to be unknown to the sender and receiver.    On the other hand,  depolarising noise in real world applications  often arises from uncontrolled fluctuations, which may occur on a short time scale. In this setting, the unitary processes taking place in each region may vary during the transmission of the system from the sender to the receiver,  especially when $N$ is large. 

\begin{figure}[htbp!]
\centering
\includegraphics[scale = 0.38]{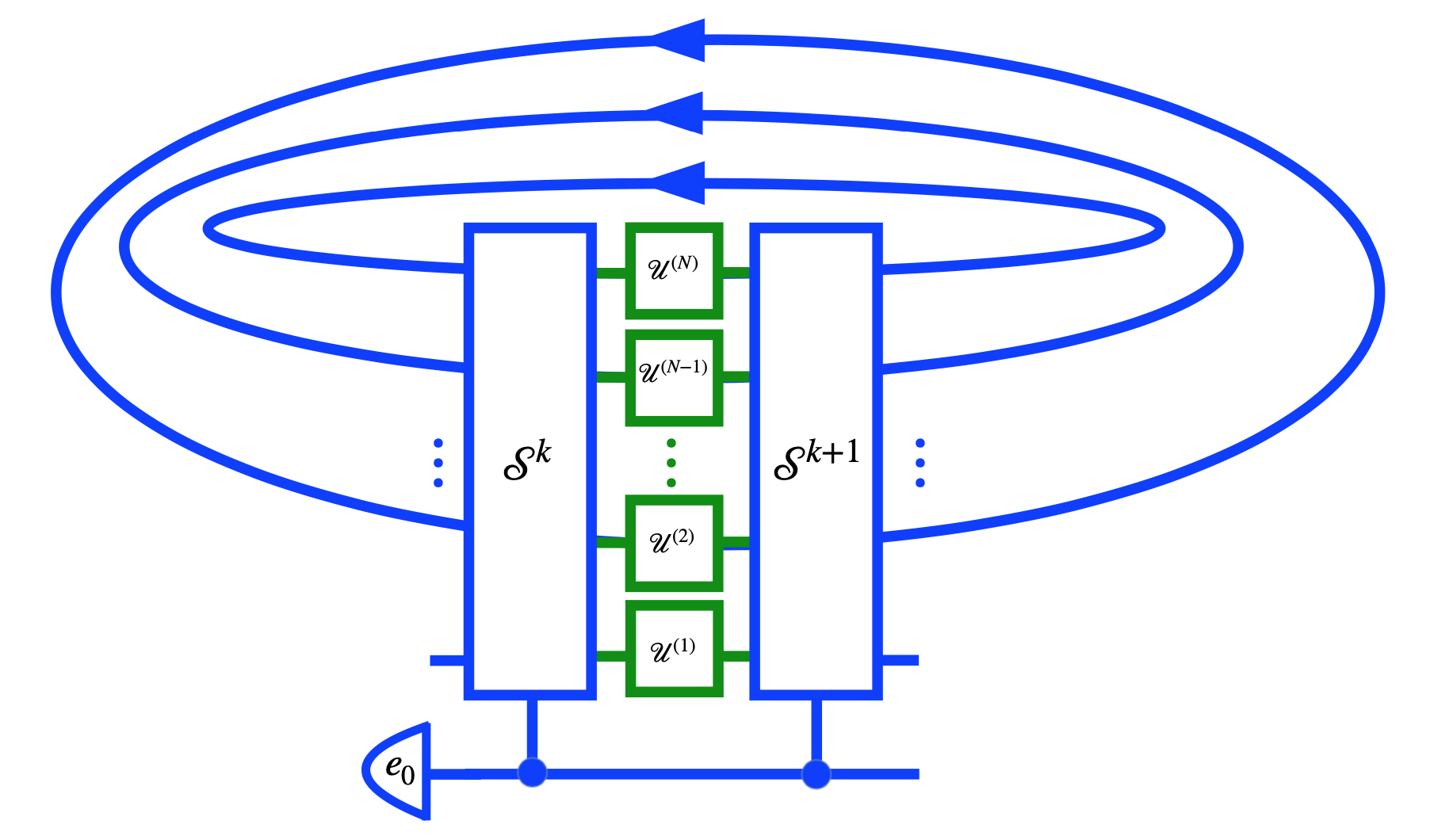}
\caption{{\bf Realisation of the effective channel $\map C_{\rm eff}$  through a circuit with loops.}   A quantum system traverses  $N$ regions via a quantum circuit including loops. The path of the system is determined by  an $N$-dimensional quantum system (bottom wire), initialised in the superposition state $|e_0\>  =  \sum_{k=0}^{N-1}  |k\>/\sqrt{N}$.  The control of the paths is implemented by unitary channels ${\rm ctrl}-\map S^{k}$ and ${\rm ctrl}-\map S^{k+1}$ corresponding to unitary operators ${\rm ctrl}-S^{k+i}:  =  \sum_{k=0}^{N-1}    S^{k+i}  \otimes  |k\>\<k|$, where $S$ is the unitary operator that shifts cyclically  $N$ quantum systems  by one position.   In the $k$-th region, the system undergoes a unitary process $\map U^{(k)}$, chosen at random according to a suitable probability  distribution. The overall evolution from the input to the output is given by the channel $\map C_{\rm eff}$ in Eq. (7) of the main text.     }
\label{fig:loops}
\end{figure}

A more radical way to generate the channel $\map C_{\rm eff}$ is to use a quantum circuit with loops, as illustrated  in Figure \ref{fig:loops}.  Mathematically, the loops in the circuit correspond to postselected teleportation protocols, where the probability of success of one of the outcomes is artificially scaled up to 1 \cite{chiribella2009beyond,lloyd2011closed}.    Physically,  circuits with loops could arise  in  scenarios involving closed timelike curves, or scenarios where postselection is taken to be fundamental \cite{oeckl2008general,svetlichny2011time,lloyd2011closed,oreshkov2015operational}.    Alternatively, they could arise probabilistically from ordinary quantum circuits using quantum teleportation. In this realisation, however,  the probability of successfully producing channel $\map C_{\rm eff}$ decreases exponentially with $N$: for $N$ qubit  channels, the success probability is $1/4^{N-1}$, due to the $N-1$ loops in the circuit.  

The realisation  in Figure \ref{fig:loops}  was presented  in  \cite{chiribella2009beyond} for $N=2$ and is generalised here to $N>2$.      An important feature   is that it does not require correlations between processes happening at different times.  Ideally, the processes $ \map U^{(k)}$ could even be taken to be instantaneous, and still the scheme in Figure \ref{fig:loops} would provide a realisation of the channel $\map C_{\rm eff}$.   This is      in stark contrast with the realisation in Figure \ref{fig:paths}, which requires each  unitary gate $\widetilde {\map U}^{(k)}$ to be the same  at $N$ different moments of time.

The scheme in Figure \ref{fig:loops} also admits a more straightforward extension from unitary to non-unitary processes: explicitly, it allows one to realise the quantum SWITCH of $N$ arbitrary channels by simply inserting such channels in the green slots of the circuit. In this respect, this scheme   reflects  more closely the definition of the quantum SWITCH as a higher order operation that takes in input $N$ quantum channels, and generates a new quantum channels by connecting them in a superposition of orders \cite{chiribella2009beyond,Chiribella2013}.


\section{Generalisation of the universal {\tt NOT} gate to $d>2$}\label{app:unot} 

The channel $\map E_1$ defined in Equation (9) of the main text is a generalisation of the universal {\tt NOT} gate to dimension $d>2$.  It can be equivalently expressed as 
\begin{align}\label{perp}
\map E_1  (\rho)  =  \int  \d \psi   \,  \psi_\perp     \,  \Tr [  P_\psi  \,  \rho] \, ,  \qquad \psi_\perp  :  =  \frac {  I  -  |\psi\>\<\psi|}{d-1}  \, ,  \quad P_\psi  :  =  d\, |\psi\>\<\psi|  \, ,   
\end{align}     
where $\d \psi$ is the unitarily invariant probability distribution over the set of pure states. 
Operationally, the channel $\map E_1$ can be realised by measuring the input system with measurement operators  $ \{ P_\psi  \}$, and preparing the output state $\psi_\perp$ conditionally on the measurement outcome $\psi$.  This implies that  $\map E_1$ is an entanglement-breaking channel \cite{horodecki2003entanglement}, and therefore cannot transmit any quantum information \cite{holevo2001evaluating}, even with the assistance  of two-way classical communication.

 Note that the channel $\map E_1$ is covariant with respect to the action of the  group $\grp {SU} (d)$, that is, it satisfies the condition 
\begin{align}
  \map U  \circ \map E_1  =   \map E_1\circ \map U \qquad     \qquad \forall \map U:  \rho  \mapsto U\rho  U^\dag \,  ,  U\in\grp{SU} (d)  \, .       
\end{align}
The covariance of $\map E_1$ follows immediately from  Eq. (\ref{perp}):  for every operator $\rho \in L(\C^d)$, one has 
\begin{align}
\nonumber 
(\map E_1  \circ \map U)( \rho)    & =  \int  \d \psi   \,  \frac {  I  -  |\psi\>\<\psi|}{d-1}      \,  \Tr [ d  (     |\psi\>\<\psi|    )    \,     U\rho U^\dag]\\
\nonumber 
 & =  \int  \d \psi   \,   U   \left( \frac {  I  -  U^\dag |\psi\>\<\psi|  U}{d-1}  \right)   U^\dag     \quad  \Tr [ d  (  U^\dag   |\psi\>\<\psi|   U )    \,     \rho ]\\
\nonumber 
 & =  \int  \d \psi'   \,   U   \left( \frac {  I  -  |\psi'\>\<\psi'| }{d-1}  \right)   U^\dag     \quad  \Tr [ d  (   |\psi'\>\<\psi'|  )    \,     \rho ]\\
&  = (\map U \circ \map E_1 )(\rho) \, , \label{covE0}
\end{align}
where the first equality follows from from  Eq. (\ref{perp}), the second equality follows from the cyclic property of the trace, the third equation follows from setting $|\psi'\>  :  =  U^\dag |\psi\rangle$ and by using the unitary invariance of $\d \psi$, and the fourth equation follows again from  Eq. (\ref{perp}). 

The channel $\map E_1$ can also be characterised as the channel for which the output state is maximally orthogonal to the input state. More precisely, $\map E_1$  minimises the fidelity between a generic input state $|\psi\>$ and the corresponding output state $\map E_1  (|\psi\>\<\psi|)$, 
 namely
\begin{align}\label{Fid}
F (\map C) :=  \int \d \psi \,   \<\psi|  \,\map C  (|\psi\>\<\psi| ) \,  |\psi\> \, .
\end{align}  

Without loss of generality, the channel $\map C$ that minimises the fidelity $F(\map C)$ can be chosen to be covariant with respect to the action of the  group $\grp {SU} (d)$. 

Note that covariant channels form a convex set. Since the fidelity (\ref{Fid}) is a linear function of $\map C$, the minimum is attained on an extreme point of the convex set of covariant channels.  
The extreme points have been classified in  Ref. \cite{chiribella2005extremal}  in terms of the Choi representation, which associates channel $\map C$  to the Choi  operator   
\begin{align}
C:  =  \sum_{i,j}    \map C  (|i\>\<j|) \otimes |i\>\<j| \,. 
\end{align} 
Specifically, Theorem 1 of Ref. \cite{chiribella2005extremal}  shows that the extreme covariant channels have Choi operators  of the form 
\begin{align}
C   =  d\,  \frac{ P}{ \Tr[P]} \, ,     
\end{align}
where $P$ is a projector on an irreducible subspace of the representation $\{  U\otimes\overline U\}_{U \in  \grp{SU}  (d)}$.  This representation has two irreducible subspaces: the first is one-dimensional and consists of vectors proportional to the maximally entangled state $|\Phi^+\>:  =\sum_i  |i\>|i\>/\sqrt{d}$, while the second is the orthogonal complement of $|\Phi^+\>$.  

The corresponding Choi operators are $C_1   =  d  \,  |\Phi^+\>\<\Phi^+|$ and $C_2  =  d/(d^2-1)   \,   (I\otimes I  -  |\Phi^+\>\<\Phi^+|)$.    Direct inspection shows that they are the Choi operators of the identity channel and of channel $\map E_1$, respectively.   The identity channel is the maximiser of the fidelity (\ref{Fid}), while channel $\map E_1$
 is the minimiser. 


\section{Application to private communication and key distribution}\label{app:applications}

In the main text we have shown that the quantum SWITCH of $N$ completely depolarising channels  (with sufficiently large $N$) enables  a heralded transmission of quantum information, with probability of success larger than $1/d^2$ and with fidelity approaching 1 as $1/N$.  The ability to open up a reliable channel for the transmission of quantum data is potentially useful for cryptographic applications, including (but not limited to) private classical communication and quantum key distribution. In the following we focus on these two applications, clarifying the underlying assumptions for the application of our results, and outlining possible ways to go beyond these assumptions.

\smallskip

{\em Private classical communication.}   The most direct cryptographic application of our results is  the transmission of private classical messages \cite{devetak2005private,horodecki2005secure}. In this task, a sender and a receiver use a  given quantum channel to communicate classical messages with the guarantee that no eavesdropper can decode the message by accessing the environment of the channel.   The rate at which classical communication can be privately transmitted is called the {\em private capacity}, and the quantum capacity of a channel is known to be a lower bound to its private capacity.  Hence, a quantum channel with non-zero quantum capacity can transmit private classical messages at a non-zero rate. This fact can be used to argue a non-zero private capacity in the context of our communication protocol, in the scenario where two-way classical communication between sender and receiver is allowed~\cite{bennett1996mixed}.  

The argument can be outlined  as follows.  The insertion of $N$ completely depolarising channels into the quantum SWITCH gives rise to the effective channel $\map C_{\rm eff}$ in Eq. (7) of the main text.  Then, a lower bound on the private capacity of $\map C_{\rm eff}$ assisted by two-way classical communication follows from two observations: 
\begin{enumerate}
\item  There exists a two-way communication protocol that uses the channel $\map C_{\rm eff}$ to achieve quantum data transmission at a rate that is at least  the average of the two-way assisted quantum capacities of channels $\map E_0$ and $\map E_1$, with the probabilities given in the main text.     The details of the protocol are presented in Section \ref{app:quantum} of this supplemental material.     
 \item Channel $\map E_0$ has non-zero quantum capacity assisted by two way communication if and only if $N\ge d +2$.  The details of the protocol are presented in Section \ref{app:zerocap1} of this supplemental material.  
  \end{enumerate}
The above results show that the effective channel $\map C_{\rm eff}$  has a non-zero quantum capacity assisted by two-way classical communication whenever $N\ge d+2$. Since the (two-way assisted) quantum capacity is a lower bound to the (two-way assisted) private capacity, this argument shows that  the quantum SWITCH of $N$  completely depolarising channels guarantees private classical communication  whenever $N\ge d+2$.

\smallskip

{\em Quantum key distribution.}  In the main text we have shown that the quantum SWITCH of $N$ completely depolarising channels  permits a perfect heralded transmission of single qubit states in the $N\to \infty $ limit.  In turn, the heralded transmission of single qubits could be used to implement the BB84 \cite{BEN84}  or E91 protocols \cite{Ekert1991QuantumTheorem} for the generation of secret keys.    An obvious question is whether this heuristic applies also in the finite $N$ scenario.  In principle, the fact that the transmission error vanishes as $1/N$ suggests an affirmative answer, as one can always find a sufficiently large finite $N$ such that the noise level in the heralded channel $\map E_0$ is below any desired error threshold.  For example,  if one uses the standard error threshold of 11\% for the BB84 protocol, then every value $N>15$ provides  a sufficiently clean communication channel.   Of course, one may want to consider more stringent thresholds that arise when the eavesdropper is allowed to perform more general types of attacks (possibly including attacks where the eavesdropper itself is allowed to perform operations in an indefinite order).  Nevertheless, as long as the threshold is finite, there would still exist a finite value of $N$ such that the quantum SWITCH provides a heralded channel above the threshold.  

 The above analysis  relies on the assumption that the  transmission line between the sender and receiver is described by a specific quantum channel, namely  the channel $\map C_{\rm eff}$  in Eq.~(7) of the main text.  To guarantee this condition,  the sender and receiver can adopt a  channel certification protocol   \cite{eisert2020quantum,wu2019efficient,wu2021efficient}, which would allow  them to decide whether or not the transmission line between them is sufficiently close to the channel $\map C_{\rm eff}$.  Recall that the effective channel  $\map C_{\rm eff}$ is a mixture of a partially depolarising channel and of a universal NOT channel, flagged by orthogonal states of the control system. In the communication protocol,  only the partially depolarising channel channel is used.
Hence, the communicating parties only need to certify a single-qubit, partially depolarising channel.  This can be done {\em e.g.}  by applying the channel to one side  of a maximally entangled state, and by certifying the resulting state with a tomographic method. Particularly suited for this purpose is the method  by Renner and Christandl \cite{christandl2012reliable}, which provides reliable error bars   even in the presence of an adversary.  Another approach, which does not require a full tomography,  is to observe that the partially depolarising channel is  covariant under the  $\grp{SU}(2)$ group.  To check whether an unknown channel is close to the desired depolarising channel, one can   apply random unitary gates at the input and output, thus converting the unknown channel into an $\grp{SU}(2)$-covariant channel,  and then use the fact that every $\grp{SU}(2)$-covariant channel is a random mixture of the identity channel and of the universal NOT (see {\em e.g.} \cite{chiribella2005extremal}). At this point, the certification of the partially depolarising channel can be achieved by estimating its fidelity with the identity channel, or equivalently,  by  estimating the  fidelity of its Choi state with the maximally entangled state \cite{zhu2019optimal}.    

Operationally, the condition that the transmission line is described by a specific channel (in our case, $\map C_{\rm eff}$)  places a constraint on the possible attacks that an eavesdropper may perform: the eavesdropper can have full access to the environment of the channel and to the classical communication between the sender and the receiver, but cannot modify the overall channel connecting them.    In the context of key distribution, however,  it is   desirable to relax this assumption.    Several relaxations are possible,   and in the following we briefly outline some of them for the benefit of researchers interested in further exploring this line of research.

A first relaxation  is  to allow the eavesdropper to alter the quantum channels placed  into the quantum SWITCH,  while still assuming that the quantum SWITCH correctly describes the way  these channels are combined.  For example, one may consider a spacetime situation where the quantum SWITCH describes the causal structure connecting the laboratories of  $N$ eavesdroppers, as in the ``closed laboratory'' model of Oreshkov, Costa, and Brukner \cite{oreshkov2012quantum}.    In this framework, the eavesdroppers are free to perform any operation in their respective laboratories, but do not interfere with the overall  causal structure.   Mathematically, this situation can be modelled with   the circuit with loops in Figure \ref{fig:loops}, assuming that the green boxes are chosen freely by $N$ eavesdroppers, while the blue portion of the circuit  is  trusted.

While the assumption that the eavesdroppers do not interfere with the superposition of causal orders may be justified in the above  spacetime scenario, it is hard to justify it ordinary realisations  based on coherent control over multiple paths, such as the realisation in Figure \ref{fig:paths} or other realisations employed in table-top implementations of quantum communication with indefinite causal order \cite{goswami2020increasing,guo2020experimental,rubino2021experimental,goswami2020experiments}.  In the  scenario of Figure \ref{fig:paths}, an  eavesdropper acting in one of the $N$ regions can  break the superposition of orders by performing measurements that determine whether or not the  region is traversed by a particle at a given time.  Note that, of course, the collapse of coherence will alter the overall channel between the communicating parties, who  can in principle detect the alteration,  and abort the protocol if necessary.  An interesting question is whether there exists a test that is weaker than a full channel certification and has the property that, if passed, it guarantees security with an arbitrary  adversarial choice of channels.     In the scheme of Figure \ref{fig:paths}, addressing this question requires  replacing  the unitary channels $\widetilde{  \map U}^{(k)}$ with arbitrary quantum channels, possibly changing over time, exhibiting correlations between one time step and the others, and/or correlations between one region and another.  
  
   Note that the model  discussed in the previous paragraph still assumes a trusted mechanism controlling the path of the system through the $N$ regions (referring to Figure \ref{fig:paths}, the assumption is  that the blue parts of the circuit are trusted).   Further relaxations of this assumption include giving the eavesdropper access to the control system, {\em e.g.} replacing the state $|e_0\>$ with an arbitrary initial state, and/or performing adversarial operations on the control system at some of the intermediate steps before the transmission is concluded.     Finally, one could give the eavesdropper complete control both over the channels and over the way they are combined.  For the communicating parties, the transmission line would then become  a black box, on which no assumption is made  except for the dimensionality of its input and output systems.    This situation is an instance of semi-device-independent cryptography, which aims at proving security under mild assumptions on the devices used in the protocol.  Key distribution protocols with bounds on the dimension  have been previously  developed  \cite{pawlowski2011semi,chaturvedi2018security} and proved to be secure against  individual attacks, possibly involving the use of quantum memories.     An interesting open problem in this research direction  is the extension of the security proofs to more general attacks.


 

\section{Quantum capacity assisted by two-way classical communication}\label{app:quantum}

Here we show that, for $N > d +1$,   the effective channel $\map C_{\rm eff}$ has a non-zero quantum capacity assisted by  two-way classical communication between sender and receiver, as in the model of Ref. 
\cite{bennett1996mixed}.  

Since both channels $\map E_0$ and $\map E_1$ are covariant with respect to the defining representation of $\grp{SU} (d)$, they can be implemented by quantum teleportation, using the Choi state as the resource state for the teleportation protocol \cite{bennett1996mixed,chiribella2009realization}. 

Hence, the sender and the receiver can use the following communication protocol:  
\begin{enumerate}
\item the sender prepares $n$  pairs of $d$-dimensional systems, with each pair in the maximally entangled state $|\Phi^+\>  = \sum_i  \, |i\> \otimes |i\>/\sqrt d$,
\item for each pair, the sender sends the first system of the pair to the receiver through the effective channel $\map C_{\rm eff}$ (all together, this means that the channel $\map C_{\rm eff}$ is used $n$ times), 
\item the receiver measures all the control systems, identifying two subsets: one subset containing of  $k$ systems acted upon by channel $\map E_0$, and another subset containing $(n-k)$ systems acted upon by channel $\map E_1$, 
\item the receiver communicates to sender which systems have been acted upon channel $\map E_0$, and which ones have been acted upon channel $\map E_1$ (at this point, the sender and receiver share $k$ copies of the Choi state of $\map E_0$, and $n-k$ copies of the Choi state of $\map E_1$, according to some suitable permutation known both to the sender and to the receiver), 
\item  the sender encodes information using the optimal encoding for two-way assisted quantum  communication with  channel $\map E_0^{\otimes k}  \otimes \map E_1^{\otimes (n-k)}$  (with suitable permutation of the systems, depending on the location of the systems acted upon by channels $\map E_0$ and $\map E_1$, respectively), 
\item the sender and the receiver use the Choi states as a resource for quantum teleportation, and thereby  achieving the transmission the encoded state through the channel $\map E_0^{\otimes k}  \otimes \map E_1^{\otimes (n-k)}$ (with suitable permutation of the systems), 
\item the sender and receiver use two-way classical communication to achieve optimal two-way assisted quantum communication with channel $\map E_0^{\otimes k}  \otimes \map E_1^{\otimes (n-k)}$ (with suitable permutation of the systems).  
\end{enumerate}
For large $n$, the above protocol achieves a rate of at least $  (1-p)   Q_{\leftrightarrow}  (\map E_0)  +  p\,  Q_{\leftrightarrow}  (\map E_1) $, where $Q_{\leftrightarrow}(\map C)$ denotes the two-way assisted quantum capacity of a generic channel $\map C$. In the particular case under consideration, we have $Q(\map E_1)  =  0$, so the contribution to the quantum capacity comes only from channel $\map E_0$.   Hence, the two-way assisted capacity of channel $\map C_{\rm eff}$ is non-zero  whenever  the two-way assisted capacity of channel  $\map E_0$ is non-zero. In the next section of this Supplemental Material, we will show that  $Q_{\leftrightarrow}  (\map E_0) $ is non-zero whenever $N$ is strictly larger than $d + 1$.

{ \section{$Q_{\leftrightarrow}  (\map E_0)  > 0$   if and only if   $N > d+1$}\label{app:zerocap1}

 The quantum channel $\map E_0$ in Eq. (8) of the main text is a partially depolarising channel, of the form $\map E_0  = \lambda_{N,d}  \, \map I  +   (1-\lambda_{N,d}) \,  \map D$ with $\lambda_{N,d}  :=  (N-1)/(N-1+d^2)$.   Its Choi operator is   
 \begin{align}
 \nonumber E_0   & = \frac 1 d \,  \sum_{m,n} \,   \map E_0  (|m\>\<n|)  \otimes |m\>\<n|  \\
  & =  \lambda_{N,d}  \,   |\Phi^+\>\<\Phi^+|    +  (1-\lambda_{N,d})  \,   \frac {  I\otimes I}{d^2}  \,, \label{isotropic}
  \end{align}   
 where $|\Phi^+\> :  =  \sum_m\,  |m\>\otimes |m\>/\sqrt{d}$ is the canonical maximally entangled state. 
 
States of the form  (\ref{isotropic}) are known as isotropic, and their entanglement properties have been studied in Refs.  \cite{horodecki1999reduction,braunstein1999separability}. In particular, these works have shown that  isotropic states of the form $\rho_\lambda  =  \lambda  \,   |\Phi^+\>\<\Phi^+|    +  (1-\lambda )  \,   \frac {  I\otimes I}{d^2}  $ are entangled if and only if $\lambda  >  1/(d +1)$.  Moreover, they showed that  isotropic states are entangled if and only if they are distillable, that is, if and only if one can extract perfect Bell states from them at a non-zero rate and and with with asymptotically negligible error.  

If the Choi state $E_0$ is separable, then the quantum channel $\map E_0$ is entanglement-breaking \cite{horodecki2003entanglement}, and therefore it has zero quantum capacity, even with the assistance of two-way classical communication.     If the Choi state $E_0$ is entangled, then it is distillable, and the distillation rate provides a lower bound to the two-way quantum capacity.  Hence, the condition for the two-way assisted quantum capacity to be positive is   $\lambda_{N,d}  >  1/(d +1)$, or equivalently, $N  >  d +1$, as it follows from the equality $\lambda_{N,d}  =  (N-1)/(N-1+d^2)$.      }

\section{No quantum information transmission with $N=2$ completely depolarising channels}\label{app:zerocap}

Here we show that the transmission of quantum information is impossible when  $N=2$ completely depolarising channels are combined in a superposition of orders.  We prove this result in three scenarios, listed in order of increasing generality: 
\begin{enumerate}
\item the two quantum channels are arranged in a superposition of two different orders, and arbitrary operations on the target system are allowed at the intermediate node between them, as in Figure 1 of the main text, with $N=2$, 
\item the two quantum channels are  arranged in a superposition of two different orders, and  controlled operations are allowed between them, 
\item the two quantum systems are arranged in a superposition of two different orders, the sender can perform operations that do not establish  entanglement between her lab and the control system,  and the intermediate party can perform arbitrary controlled operations.
\end{enumerate}

The impossibility of transmitting quantum information in these three cases is demonstrated in the following subsections.

\subsection{Intermediate party without access to the control}
 
 Suppose that two completely depolarising channels are placed in a superposition of two alternative orders, and that the intermediate party acting between them performs an operation described by a quantum channel $\map R$.  
 
Mathematically, the two depolarising channels are first combined by the (original) quantum {\tt SWITCH}, thus obtaining the channel 
\begin{align}
\map S  [\mathcal{C}^{(1)} ,  \mathcal{C}^{(2)}]  (\cdot)  =  \sum_{j_1,  j_2}    W_{j_1 j_2}  \cdot   W_{j_1j_2}^\dag  \, ,
\end{align}
with  Kraus operators $W_{ij}$ 
\begin{align}
W_{j_1j_2}   :  =  |0\>\<0| \otimes C^{(1)}_{j_1}  \otimes  C^{(2)}_{j_2}      +|1\>\<1| \otimes     C^{(2)}_{j_2}  \otimes  C^{(1)}_{j_1}  \, .   
\end{align} 
Then,  the quantum channel $\map R$ is inserted between the first and second time-slot, resulting into a new channel with Kraus operators   
\begin{align}
W_{j_1j_2 k}'   :  =  |0\>\<0| \otimes C^{(1)}_{j_1}  R_k C^{(2)}_{j_2}      +|1\>\<1| \otimes     C^{(2)}_{j_2}  R_k  C^{(1)}_{j_1}  \, ,  
\end{align} 
where $\{R_k\}$ are the Kraus operators of channel $\map R$.

When the control is initialised in the state $\omega$, the  resulting channel  $\map C_{{\rm eff},\omega, \map R}$ can be computed using Equations (\ref{iduno}) and (\ref{idue}), and is given by 
\begin{align}
\map C_{{\rm eff}, \omega, \map R}   (\rho)   :  =  \omega_{00}     \,  |0\>\<0| \otimes \frac I d   +  \omega_{11} \, |1\>\<1| \otimes \frac I d    +\Big(  \omega_{01} \, |0\>\<1|  +  \omega_{10}\,  |1\>\<0|  \Big)   \otimes \frac{ \map R^\dag (\rho)}{d^2} \, ,
\end{align}
where $\map R^\dag  $ is the adjoint of $\map R$, defined as $\map R^\dag  (\rho)   :  =  \sum_k R_k^\dag \rho  R_k$ where $\map R (\rho)  =  \sum_k R_k \rho R_k^\dag$ is an arbitrary Kraus decomposition of $\map R$ (the definition of $\map R^\dag$ is independent of the choice of Kraus decomposition). 

Since  the channel $\map R$ satisfies the normalisation condition $\sum_k R_k^\dag R_k =  I$,  the above relation can be rewritten as 
\begin{align}\label{ceffomegar}
\map C_{{\rm eff}, \omega, \map R}    =   (\map I_C  \otimes \map R^\dag)  \circ \map C_{\rm eff,\omega}  \, ,  
\end{align}
where  $\map I_C$ is the identity on the control system, and $\map C_{\rm eff ,\omega} : =   \map C_{{\rm eff},\omega, \map I}$. 

Now, the action of the channel $\map C_{\rm eff,\omega}$ can be expressed as  
\begin{align}\label{ceffomega}
\map C_{\rm eff,\omega}  (\rho) = \map J_{\rm eff}  (\omega \otimes \rho)  
\end{align}
where $\map J$ is a quantum channel acting jointly on the control and the target, given by 
\begin{align}\label{Jeff}
\map J_{\rm eff} ( \omega \otimes \rho) : =   p_+  \, \omega \otimes  \map E_+   (\rho)       + p_-  \, Z\omega Z \otimes   \map E_- (\rho)   \, ,
\end{align} 
with   $Z =  |0\>\<0|  -  |1\>\<1|$,   $p_\pm  :=    (d^2\pm 1)/(2d^2)$,  and \begin{align}\label{Epm}
\map E_\pm (\rho)   :=  \frac{1}{   d^2\pm1} \,  \left  (  d  \,I   \pm   \rho  \right)
 \, .
\end{align}  
 Note that the channels   $\map E_+$ and $\map E_-$ coincide with the channels $\map E_0$ and $\map E_1$ in the main text, in the special case $N=2$.  
 As it turns out, both channels $\map E_+$ and $\map E_-$ are entanglement-breaking \cite{horodecki2003entanglement}, and therefore cannot be used to transmit quantum information.   For $\map E_+$, the proof is provided in Appendix \ref{app:zerocap1}, while for $\map E_-$ the proof is provided in Appendix \ref{app:unot}.  
 Hence, also the channel $\map J_{\rm eff}$ in Equation (\ref{Jeff}) is entanglement-breaking, and so are the channels $\map C_{\rm eff, \omega}$  and  $\map C_{{\rm eff}, \omega ,\map R}$ in Equation (\ref{ceffomegar}).  

\subsection{Intermediate party performing controlled operations} 
 Suppose that two completely depolarising channels are placed in a superposition of two alternative orders, and that the intermediate party acting between them can perform controlled operations, of the form
 \begin{align}
 {\tt ctrl-}\map R (\omega \otimes \rho)  :  = \sum_k  \,  \left(   |0\>\<0|  \otimes  R_k  + |1\>\<1|  \otimes  R'_k    \right)\,   (\omega \otimes \rho)  \,  \left(   |0\>\<0|  \otimes  R_k  + |1\>\<1|  \otimes  R'_k    \right)^\dag  \, , \end{align} 
 with $\sum_k R_k^\dag R_k  =  \sum_k R_k^{\prime \dag}  R_k'  =  I$.  

When the control is initialised in the state $\omega$, the effective channel is   
\begin{align}
\nonumber \map C_{{\rm eff},\omega, {\tt ctrl-}\map R}   (\rho)     &  =  \omega_{00}     \,  |0\>\<0| \otimes \frac I d    +  \omega_{11} \, |1\>\<1| \otimes \frac I d   +   \omega_{01} \, |0\>\<1|   \otimes \left( \frac{\sum_k  R_k^\dag \rho R_k'}{d^2}  \right) +  \omega_{10}\,  |1\>\<0|   \otimes \left( \frac{\sum_k  R_k^{\prime \dag} \rho R_k  }{d^2}\right) \\
& =   {\tt ctrl-}\map R^{\dag}    \circ \map C_{\rm eff ,\omega}  (\rho ) \, .
\end{align}
In the previous subsection we showed that  $\map C_{\rm eff,\omega}$ is entanglement-breaking. Hence, $ \map C_{{\rm eff}, \omega, {\tt ctrl-}\map R}$ is entanglement breaking as well.   

\subsection{Sender performing operations that do not create entanglement between her lab and the control}

We now consider the scenario where  two completely depolarising channels are placed in a superposition of two alternative orders, and all the parties involved in the communication protocols (sender, receiver, and intermediate party acting between the two depolarising channels) can perform joint  operations on the control system, subject to the constraints that {\em (i)} the sender cannot establish entanglement between her laboratory and the control system, and {\em (ii)} the intermediate party performs a controlled operation.

Let us denote by  
 $\map A, {\tt ctrl-}\map R, $ and $\map B$ the channels performed by sender, intermediate party, and receiver, respectively. The  condition that the sender does not  use the control to establish entanglement is captured by the following definition: 
 \begin{defi}
 A channel $\map A$ from an input system $C\otimes S$ to an output system  $C'\otimes S'$ {\em does not transfer  entanglement from  $S$ to $C'$} if, for every initial state $\omega$ of system $C$, and every entanglement-breaking channel $\map E$ acting on system $S'$, the channel $(\map  I_{C'} \otimes \map E) \circ \map A  (\omega \otimes \map I_S)$ is entanglement-breaking.
 \end{defi} 
 Intuitively, the definition means that, once the entanglement with system $S'$ is destroyed, no entanglement remains in the output. 

The effective channel resulting from the superposition of orders and from the parties' operations is 
 \begin{align}   
 \map C_{{\rm eff}, \omega, \map A, \map R ,\map B}  (\rho )   = \map B \circ {\tt ctrl-}\map R^\dag \circ   \map J_{\rm eff}  \circ \map A  \circ (\map I_C\otimes \omega)\, , 
 \end{align}
where $\map J_{\rm eff}$ is the channel defined in Equation (\ref{Jeff}).  Note that one has 
\begin{align}\label{poi}
\map J_{\rm eff}  \circ \map A  \circ (\map I_C\otimes \omega)  = p_+  \,     (\map I_C \otimes \map E_+) \circ  \map A  \circ (\map I_C\otimes \omega)    +  p_-   \,   (\map Z \otimes \map I) \circ (\map I_C \otimes \map E_-)  \circ \map A \circ (\map I_C\otimes \omega)  \, , 
\end{align}
where $\map Z$ is the unitary channel defined by $\map Z (\rho)  :  = Z \rho Z$. 
Now, recall that both $\map E_+$ and $\map E_-$ are entanglement-breaking. Using the fact that $\map A$ does not transfer entanglement to the control, we obtain that the two terms in the r.h.s. of Eq.~(\ref{poi})   are entanglement-breaking channels. Hence,  the whole channel $\map J_{\rm eff}  \circ \map A  \circ (\map I_C\otimes \omega)  $ is entanglement-breaking, and so is the effective channel  $ \map C_{{\rm eff}, \omega, \map A, \map R ,\map B}$.  In conclusion, no transmission of quantum information is possible unless the sender transfers entanglement to the control.

{
\section{Partially depolarising channels in the $N=2$ scenario}\label{app:partial}  

We have shown that the combining     $N=2$ {\em completely} depolarising channels  in the quantum SWITCH does not enable a transmission of quantum data.   An interesting question is whether the quantum SWITCH could still  permit  quantum data transmission  using  $N=2$ {\em partially} depolarising channels that, individually, have zero quantum capacity.  Here we answer the question in the negative, showing that the quantum capacity of each partially depolarising channel  is a bottleneck for  quantum capacity achievable  through the quantum SWITCH (even in a heralded setting). 
In particular, this result implies that no information can be sent through the quantum SWITCH unless the original depolarising channels already  had a positive capacity.


Consider the partially depolarising channel $\map D_\lambda  :  =  \lambda\,  \map I+    (1-\lambda) \,  \map D$, where $\lambda$ is the probability that no depolarisation takes place, and $\map D$ is the completely depolarising channel.     Suppose that two identical channels $\map C^{(1)}  =  \map C^{(2)}    = \map D_p $ are placed in the quantum SWITCH, with the control qubit initially in the state $|+\>$.    Then, the resulting channel is   
\begin{align}  \map S  (\map C^{(1)},  \map C^{(2)})    =  \lambda^2\,   |+\>\<+|    \otimes   \,  \map I     + 2 \lambda  (1-\lambda) \,  |+\>\<+|       \otimes   \map D     +  (1-\lambda)^2  \,      (   p_0\,  |+\>\<+|     \otimes \map E_0   + p_1  \,   |-\>\<-|    \otimes   \map E_1   ) \, ,   
\end{align}
where   $\map E_0   =  \frac{  1  }{   d^2+1}  \, \map I +   \frac{d^2}{   d^2 +1} \,   \map D$, and $\map E_1    =  \frac{d^2}{  d^2-1  }\,  \map  D    -   \frac 1 {d^2 -1} \, \map I$ are the quantum channels defined in Eqs. (8) and (9)  of the main text,   while $p_0  = (d^2+1)/(2d^2) $ and $p_1    =  (d^2-1)/(2d^2) $ are the corresponding probabilities  (cf. Eq. (7) in the main text).  

The channel $\map S  (\map C^{(1)},  \map C^{(2)})$ can also be written as 
\begin{align}
\map S  (\map C^{(1)},  \map C^{(2)})    =  [1-  (1-\lambda)^2 \, p_1]   \,   |+\>\<+|    \otimes    \map E_0' + (1-\lambda)^2 \, p_1  \,   |-\>\<-|    \otimes   \map E_1     \qquad    \map E_0'  :  =  \frac{    p^2   \,  \map I     +  2 \lambda  (1-\lambda) \, \map D     +  (1-\lambda)^2  \,       p_0\,   \map E_0 }{1-(1-\lambda)^2 \, p_1}   \, .  
\end{align}
As in the fully depolarising case,  it is  possible to separate the channel $\map E_0'$ and $\map E_1$ by measuring the order qubit on the basis $\{ |+\>, |-\>\}$.  When such a measurement is performed, the outcome $+$ heralds the occurrence of the channel  $\map E_0'$, while the outcome $-$ heralds the occurrence of the channel    $\map E_1$.

Now,  recall that the channel $\map E_1$ is entanglement breaking, and therefore cannot transmit any quantum information. Hence, we will focus our attention on the channel $\map E_0'$, heralded by the outcome $+$.     The  heralded channel $\map E_0'$ is a partially depolarising channel,  of the form $\map E_0'    =  \lambda' \, \map I  + (1-\lambda')\,  \map D $,  with 
\begin{align}\label{pprime} \lambda'  =  \frac{  \lambda^2      +    \,     \frac{  (1-\lambda)^2  }{2 d^2}     }{1-\frac{(1-\lambda)^2 \,   (d^2-1)}{2d^2} }    \, .
\end{align}

We now show that  the heralded channel   $\map E_0'$  has less   quantum capacity than a single depolarising channel $ \map D_\lambda$.  In other words, the quantum capacity of the each individual  depolarising channel remains a bottleneck in the $N=2$ case.    To establish this result, we compare the new probability $\lambda'$ in Eq. (\ref{pprime}) with the probability $\lambda$ appearing in the original depolarising channels $ \map D_\lambda$.   
By solving the inequality $\lambda'  \ge \lambda$,   we obtain  the solutions $\lambda=1$ and  $\lambda \le   \frac{d  \sqrt{d^2  +8  }  -  d^2   -2}{2 (d^2  -1)} $.  

A first observation  is that the quantum SWITCH  does not break the  $ \map D_\lambda$  bottleneck   in the low noise regime. For $\lambda  \approx 1$,  one has $\lambda'<\lambda$, meaning that, in fact, the heralded channel $\map E_0'$ is more noisy than the original depolarising channel  $\map D_\lambda$.

Let us now look into the high noise regime  $\lambda \le   \frac{d  \sqrt{d^2  +8  }  -  d^2   -2}{2 (d^2  -1)}  =: \lambda_{\max} $.     Recall that a depolarising channel $\map D_\lambda$ is entanglement breaking whenever $\lambda   \le 1/(d+1)$ (see {\em e.g.}  Supplementary Note  \ref{app:zerocap1}).    As it turns out, one has    $\lambda_{\max}  <  1/(d+1)$, meaning that the quantum SWITCH reduces the amount of noise only when the initial channel $\map D_\lambda$ is entanglement breaking.      In this regime, however, also the heralded channel $\map E_0'$ is also entanglement breaking. Summarising, the quantum SWITCH of $N=2$   depolarising channels reduces the amount of noise  for $\lambda  \le \lambda_{\max}$,   but this noise reduction is not sufficient to enable the transmission of quantum data: if the depolarising channel  $ \map D_\lambda$ has zero quantum capacity, then  the heralded  channel $\map E_0'$ is entanglement breaking and therefore cannot transmit any quantum information.   In general,  the quantum capacity of the channel  $ \map D_\lambda$ remains as a bottleneck for the quantum capacity of the heralded channel $\map E_0'$.

In passing, we note that, while  the heralded  channel $\map E_0'$ cannot offer any advantage over a single depolarising channel $\map D_\lambda$, it still offers an advantage over  the channel $\map D_\lambda^2$, arising from the use of the two depolarising channels $\map C^{(1)}$ and $\map C^{(2)}$    in a fixed order  without any intermediate operation between them.  Since  $\map D_\lambda^2 $ is a depolarising channel of the form $\map D_\lambda^2  =  \lambda^2 \, \map I  +  (1-\lambda^2)  \, \map  D$,   we can observe that the heralded channel $\map E_0'$ is generally less noisy than   $\map D_\lambda^2$.  Indeed, the condition  $\lambda'  \ge   \lambda^2$ is satisfied   for every $p  \in  [0,1]$.  This condition implies that   channel $\map E_0'$ achieves quantum data transmission for  larger values of $\lambda$ compared to channel    $\map D_\lambda^2$.    
Note that channel   $\map D_\lambda^2$ can be obtained from the switched channel  $\map S  (\map C^{(1)},  \map C^{(2)})$ by  decohering the order qubit in the basis $\{  |0\> ,|  1\>\}$.   In this respect,  the advantage of the channel  $\map E_0'$ compared to the channel   $\map D_\lambda^2$ can be interpreted as a benefit of quantum coherence in the order qubit.

}

\section{Classical capacity of the effective channel}\label{app:capswitch}

Here we determine the classical capacity of the effective channel $\map C_{\rm eff}$ defined in Equation (3) of the main text. 
According to the Holevo-Schumacher-Westmoreland theorem \cite{holevo1998capacity,schumacher1997sending}, 
the classical capacity of a generic quantum channel $\map C$ is 
\begin{align}  
C(\map C)  =  \liminf_{n\to \infty}\,  \frac{\chi  (\map C^{\otimes n})}n  \, ,
\end{align}
where $\chi  (\map D)$ is the Holevo information of a generic quantum channel $\map D$, and is defined as $\chi(\map D)  :  =  \sup_{  \{  \rho_x  , p_x\}}   \,  S (  \sum_x p_x  \, \rho_x)   - \sum_x  p_x  \,  S(\rho_x)$, the maximum being over all ensembles $\{  \rho_x  \,  ,p_x\}$ where $\rho_x$ is a quantum state and $p_x$ is a probability.  

Now, the effective channel has the form  $\map C_{\rm eff}  =  p\,  \rho_0 \otimes \map E_0  + (1-p) \,  \rho_1\otimes \map E_1$, where  $\rho_0$ and $\rho_1$ are orthogonal states of the control system, and $\map E_0$ and $\map E_1$ are two channels acting only on the target system.  
Hence, the $n$-fold product  $\map C_{\rm eff}^{\otimes n}$ has the form 
\begin{align}
\map C_{\rm eff}^{\otimes n}   = \sum_i \,  p_i   \,  \rho_i^{(n)}\otimes  \map E_i^{(n)}  \, , 
\end{align}
  where  $\{\rho_i^{(n)}\}$ are orthogonal states of the control system, and $\map E_i^{(n)}$ is a channel acting on $n$ copies of the target system (specifically,  each state $\rho_i^{(n)}$ is the tensor product of $k$ copies of the state $\rho_0$ and $(n-k)$ copies of the state $\rho_1$, while each  channel $\map E_i^{(n)}$ is the tensor product of $k$ copies of channel $\map E_0$ and $(n-k)$ copies of the channel $\map E_1$, for  some $k \in \{0,\dots, n\}$).  
  
 By convexity of the Holevo information,   one has the inequality 
 \begin{align}
 \nonumber \chi(\map C_{\rm eff}^{\otimes n})  &  \le  \sum_i  \,  p_i^{(n)}   \,\chi  \left(   \rho_i^{(n)} \otimes \map E_i^{(n)}\right) \\
   &  =  \sum_i  \,  p_i^{(n)}   \,\chi  \left(   \map E_i^{(n)}\right)   \, ,
  \end{align}
  the second equality being due to the fact that the state $\rho_i^{(n)}$ is independent of the input of the channel. Note that, at this point, the state of the control system has disappeared from our upper bound.  In fact, the only role of the control system is to guarantee the achievability of the upper bound: since the states $\rho_i^{(n)}$ are orthogonal, it turns out that the above inequality is actually an equality.

 At this point, we observe that each channel $\map E_i^{(n)}$ is covariant with respect to the  defining  representation of the group $\grp {SU}  (d)^{\times n}$, namely  
 \begin{align}
 \map E_i^{(n)}  \circ (\map U_1\otimes \cdots \otimes \map U_n)   =  (\map U_1\otimes \cdots \otimes \map U_n) \circ  \map E_i^{(n)} \,,   
 \end{align} 
where, for every $i\in \{1,\dots,  n\}$,  $\map U_i$ is a unitary channel defined by  $\map U_i (\cdot) =   U_i  \cdot U_i^\dag$, and  $U_i$ is an arbitrary element of $ \grp{SU}(d)$, chosen independently for every value of $i$. 
Covariance of the channel $\map E_i^{(n)}$  with respect to the representation $\{  U_1\otimes \cdots \otimes U_n\}$ is derived  from the followings observations 
\begin{itemize}
\item each channel $\map E_i^{(n)}$ has  the product  form $\map E_i^{(n)}   =  \map C_1 \otimes \map C_2 \otimes \cdots \otimes \map C_n$, where each channel in the product on the right-hand-side is either the channel $\map E_0$ or the channel $\map E_1$
\item   The channels $\map E_0$ and $\map E_1$ are both covariant with respect to the fundamental representation of $\grp{SU} (d)$, that is $\map U\circ \map E_i  =  \map E_i \circ \map U$ for every  $i\in  \{0,1\}$, and for every $\map U:  \rho  \mapsto  U \rho U^\dag$, $U \in \grp{SU} (d)$.  Indeed, $\map E_0$ was proven to be covariant in Eq. (\ref{covE0}), and  $\map E_1$ is a depolarizing channel, whose covariance is immediate from the definition. 
\item Since each channel in the product $ \map C_1 \otimes \map C_2 \otimes \cdots \otimes \map C_n$ is covariant, the product is also covariant: for arbitrary and independently chosen unitary operators $U_1,U_2, \dots, U_n$, one has $(\map U_1\otimes \cdots \otimes \map U_n) \circ (\map C_1 \otimes \map C_2 \otimes \cdots \otimes \map C_n) =  (\map U_1 \circ \map C_1 )\otimes (\map U_2 \circ \map C_2 )\otimes \cdots \otimes (\map U_n \circ\map C_n)   =   (\map C_1 \circ \map U_1 )\otimes (\map C_2 \circ \map U_2 )\otimes \cdots \otimes (\map C_n \circ\map U_n)  = (\map C_1 \otimes \map C_2 \otimes \cdots \otimes \map C_n)  \circ  (\map U_1\otimes \cdots \otimes \map U_n)$. 
\end{itemize}
Note that the representation $\{  U_1\otimes \cdots \otimes U_n\}$ is irreducible.  For a generic channel $\map C$, if $\map C$ is covariant with respect to an irreducible representation, then the Holevo information has the form 
\begin{align}
\chi (\map C)   =  \log d   -  S_{\min}  (\map C)  \, ,
\end{align}
where $S_{\min} (\map C)  :  = \min_{\rho}   S  ( \map C (\rho))$ is the minimum output entropy \cite{holevo2002remarks}.  
Hence, the channels $\map E_i^{(n)}$ satisfy the condition 
\begin{align}
\chi \left(\map E_i^{(n)}\right)   =    n  \log d   -  S_{\min}  \left(\map E_i^{(n)}\right)  \, .
\end{align}

Now, it only remains to determine the minimum output entropy of the channels $\map E_i^{(n)}$. Each channel $\map E_i^{(n)}$ is the product of  $n$ channels, each of which is one of the two channels $\map E_0$ and $\map E_1$ in the main text. 

Channel $\map E_0$ is a depolarising channel, and the minimum output entropy  $S_{\min}  (\map E_0^{\otimes n})$ has been evaluated by King \cite{king2003capacity}, 
who showed the additivity property 
\begin{align}
S_{\min}  (\map E_0^{\otimes n})    =  n  S_{\min}  (\map E_0) \, ,
\end{align}  
and evaluated the minimum output entropy
\begin{align}
S_{\min}  (\map E_0)     =     - \left(  \lambda_{N,d}  +  \frac{1-\lambda_{N,d}} d \right)  \, \log  \left(  \lambda_{N,d}  +  \frac {1-\lambda_{N,d}} d\right)    -   \frac{(d-1)\, (1-\lambda_{N,d})  }d   \,  \log  \left( \frac{1-\lambda_{N,d}}d  \right)\, , 
\end{align}
where $\lambda_{N,d}$ is the probability of the identity channel  in the decomposition $\map E_0  (\rho)  =  \lambda_{N,d}  \, \rho  +  (1-\lambda_{N,d})   \,  I/d $. 

Channel $\map E_1$, the universal {\tt NOT} gate, transforms the input state $\rho$ into the output state  $\map E_1   (\rho)=   d^2/(d^2-1)  \, I/d   -  \rho/(d^2-1)$. Decomposing the input state as   $\rho  =  \sum_i \,   q_i   \,  |\psi_i\>\<\psi_i|$ for some probabilities $\{q_i\}$, and using the concavity of the von Neumann entropy, one obtains 
\begin{align}
\nonumber S(\map E_1(\rho))   & \ge \sum_i  \, q_i  \,   S(\map E_1  (|\psi_i\>\<\psi_|)) \\
   &  = S  (\map E_1  (|\psi\>\<\psi|)  \qquad \forall |\psi\>  \in  \spc H  \,  , \|  |\psi\>  \|  = 1 \, .
\end{align}
Hence, the minimum output entropy of $\map E_1$  is given by 
\begin{align}
\nonumber S_{\min}  (\map E_1)  &=   S  (\map E_1  (|\psi\>\<\psi|)  \qquad \forall |\psi\>  \in  \spc H  \,  , \|  |\psi\>  \|  = 1 \\
&   =  -\frac 1 {d+1}  \log  \left(\frac 1{d+1}  \right) -  \frac d {d+1}   \log \left( \frac{  d}{d^2-1}  \right)
   \, .  
\end{align}

Now, recall that $\map E_1$ is entanglement-breaking.  Shor  \cite{shor2002additivity}  
showed the additivity property  
\begin{align}
S_{\min}  (\map A\otimes \map B)   =  S_{\min}  (\map A)  +  S_{\min}  (\map B) \, ,
\end{align}
for every  pair of channels $(\map A, \map B)$ such that at least one of the channels is entanglement breaking.  
Hence, every channel $\map E_i^{(n)}$ of the form $\map E_i^{(n)}  =  \map E_0^{\otimes k} \otimes   \map E_1^{\otimes (n-k)}$, $k<n$,  up to permutations of the Hilbert spaces, will satisfy the condition 
\begin{align}
\nonumber S_{\min}  \left(  \map E_i^{(n)} \right)  &=    S_{\min}   (\map E_0^{\otimes k})    +  S_{\min}   (\map E_1^{\otimes  (n-k)})\\
 &=   k\,  S_{\min}   (\map E_0)    + (n-k) \,  S_{\min}   (\map E_1)   \, .
\end{align}

Hence, we obtained the bound 
\begin{align}
\nonumber \chi(\map C_{\rm eff}^{\otimes n})    & \le    \sum_{k=0}^{n}   (  1-p)^{k} p^{n-k}   \begin{pmatrix}   n\\k\end{pmatrix} [  n\log d  -   k \, S_{\min}  (\map E_0)  - (  n-k) \,  S_{\min}  (\map E_1) ]\\
  &=     n \,  \left[  \log d   -  (1-p)  S_{\min}  (\map E_0)   -  p \,  S_{\min}  (\map E_1) \right]     \, ,
\end{align} 
and therefore 
\begin{align}\label{capacitybound}
C(\map C_{\rm eff})   \le  \log d   -  (1-p)  \,S_{\min}  (\map E_0)   -  p\,  S_{\min}  (\map E_1) \, .
\end{align}

The above bound is achievable by the ensemble of orthogonal states  $\{   |x\>\<x|  \}_{x=1}^d$ with uniform probabilities $p_x = 1/d$.    Using the definition of $\map C_{\rm eff}$ in the main text, we have
\begin{align}
S \left( \map C_{\rm eff}  (|x\>\<x|)\right) =     (1-p)   \,S_{\min}  (\map E_0)   +  p   \, S_{\min}  (\map E_1)   +  H(p)   +  p \, \log (N-1) \qquad \forall x   \in \{1,\dots, d\} \, ,
\end{align}
 with $H(p)  :  =  - p\log p  -  (1-p) \log (1-p)$. 
 Moreover, we have 
\begin{align}
\map C_{\rm eff}    \left(\sum_x  \, p_x  \,|x\>\<x|  \right)   =   \frac {  I }d \otimes   \Big(  (1-p)  \, \rho_0    +   p\,   \rho_1  \Big)  \, ,  
\end{align}
from which we obtain 
 \begin{align}
S \left( \map C_{\rm eff}  \left(  \sum_x  \,  p_x \, |x\>\<x|\right)\right) =     \log d   +  H(p)   +  p\, \log(N-1) \, .
\end{align}

Hence, we have 
\begin{align}
\nonumber \chi  (  \map C_{\rm eff})  &\ge  S \left( \map C_{\rm eff}  \left(  \sum_x  \,  p_x \, |x\>\<x|\right)\right)   - \sum_x  \, p_x  S \left( \map C_{\rm eff}  (|x\>\<x|)\right)   \\
\label{chibound} &  =  \log d  -   (1-p)  \,S_{\min}  (\map E_0)   -  p  \, S_{\min}  (\map E_1)   \, .
\end{align}

Since  one has $\chi  (  \map C)  \le C (\map C)$ for every channel $\map C$,   the bounds (\ref{capacitybound}) and(\ref{chibound}) imply the equality  
 \begin{align}
 \nonumber C(\map C_{\rm eff})    &= \log d   -  (1-p)  \,S_{\min}  (\map E_0)   -  p \,  S_{\min}  (\map E_1)\\
\nonumber  &  =  \log d    +\frac {  N+d^2-1}{Nd^2}  \, \left[   \frac {  N-1  +d} {N-1+d^2}  \log \left ( \frac {  N-1  +d} {N-1+d^2} \right)    +    \frac{d \,(d-1) }{N-1 + d^2}   \,  \log  \left( \frac{  d}{N-1+d^2}\right) \right] \\
 \nonumber  & \quad   +  \frac{(N-1) (d^2-1)}{Nd^2}  \left[    \frac 1 {d+1}  \log \left( \frac 1 {d+1} \right)   +  \frac d {d+1}   \log \left( \frac{  d}{d^2-1}  \right) \right]   \\
  \nonumber  &  =  \log d    +   \frac {  N-1  +d} {Nd^2}  \log \left ( \frac {  N-1  +d} {N-1+d^2} \right)    +    \frac{d \,(d-1) }{N d^2}   \,  \log  \left( \frac{  d}{N-1+d^2}\right)  \\
  & \quad   +  \frac{(N-1) (d^2-1)}{Nd^2}  \left[    \frac 1 {d+1}  \log\left( \frac 1 {d+1} \right)   +  \frac d {d+1}   \log \left( \frac{  d}{d^2-1}  \right) \right]   \, .  \label{capacity}
\end{align}

For every fixed $N$,  the capacity vanishes as $O(1/d^2)$ in the large $d$ limit.   
In the large $N$ limit, the channel $\map E_0$ becomes noiseless, and the  capacity  (\ref{capacity})  has the asymptotic expression  
\begin{align}
C(\map C_{\rm eff})    &   =  \frac{\log  (d+1)}{d^2}   -  \left(1  - \frac 1{d}\right)  \,  \log \left( 1 -  \frac 1{d^2}\right) -  \frac 1d \log  \left (1  + \frac 1d \right)    -   O   \left(   \left (  1-\frac 1d\right)  \,  \frac {\log N} N\right) \, .
\end{align}
This expression is  decreasing with $d$, and  also converges to 0 for $d \to \infty$.   



\end{widetext}

\end{document}